\documentclass[aps,prd,superscriptaddress,twocolumn,a4paper]{revtex4-2}
\usepackage[colorlinks=true, pdfstartview=FitV, linkcolor=blue, citecolor=red, urlcolor=black]{hyperref}
\usepackage{graphicx}
\usepackage[all]{xy}
\usepackage{amsmath}
\usepackage{amssymb}
\usepackage{tensor}
\usepackage{comment}
\usepackage{orcidlink}
\usepackage[normalem]{ulem}

\newcommand{\be}{\begin{equation}}
\newcommand{\ee}{\end{equation}}
\newcommand{\ben}{\begin{eqnarray}}
\newcommand{\een}{\end{eqnarray}}
\newcommand{\bes}{\begin{subequations}}
\newcommand{\ees}{\end{subequations}}
\def\bal#1\eal{\begin{align}#1\end{align}}

\newcommand{\nn}{\nonumber\\}
\newcommand{\bfi}{\begin{figure}}
\newcommand{\efi}{\end{figure}}
\newcommand{\bc}{\begin{center}}
\newcommand{\ec}{\end{center}}

\newcommand{\sech}{\mbox{sech}}

\newcommand{\LL}{{\cal L}}
\newcommand{\Sc}{{\cal S}}
\newcommand{\Nc}{{\cal N}}

\newcommand{\Ac}{{\cal A}}
\newcommand{\Fc}{{\cal F}}

\newcommand{\Ec}{{\cal E}}
\newcommand{\Qc}{{\cal Q}}
\newcommand{\Uc}{{\cal U}}

\newcommand{\n}{\nabla}

\newcommand{\vphi}{\varphi}

\begin{document}
\title{Analytical solutions for Maxwell-scalar system on radially symmetric spacetimes}
\author{I. Andrade\,\orcidlink{0000-0002-9790-684X}}
        \email[]{andradesigor0@gmail.com}\affiliation{Departamento de F\'\i sica, Universidade Federal da Para\'\i ba, 58051-970 Jo\~ao Pessoa, PB, Brazil}
\author{D. Bazeia\,\orcidlink{0000-0003-1335-3705}}\email[]{bazeia@fisica.ufpb.br}
    \affiliation{Departamento de F\'\i sica, Universidade Federal da Para\'\i ba, 58051-970 Jo\~ao Pessoa, PB, Brazil}
\author{M.A. Marques\,\orcidlink{0000-0001-7022-5502}}
        \email[]{marques@cbiotec.ufpb.br}\affiliation{Departamento de Biotecnologia, Universidade Federal da Para\'\i ba, 58051-900 Jo\~ao Pessoa, PB, Brazil}
\author{R. Menezes\,\orcidlink{0000-0002-9586-4308}}
     \email[]{rmenezes@dcx.ufpb.br}\affiliation{Departamento de Ci\^encias Exatas, Universidade Federal
da Para\'{\i}ba, 58297-000 Rio Tinto, PB, Brazil}\affiliation{Departamento de F\'{\i}sica, Universidade Federal de Campina Grande,  58109-970 Campina Grande, PB, Brazil}
\author{G.J. Olmo\,\orcidlink{0000-0001-9857-0412}}
        \email[]{gonzalo.olmo@uv.es}\affiliation{Departament de F\'isica Te\`orica and IFIC, Centro Mixto Universitat de
Val\`{e}ncia - CSIC. Universitat de Val\`{e}ncia, Burjassot-46100, Valencia, Spain}
\affiliation{Universidade Federal do Cear\'{a} (UFC), Departamento de F\'{i}sica, Campus do Pici, Fortaleza, CE, 60455-760, Brazil}

\begin{abstract}
We investigate Maxwell-scalar models on radially symmetric spacetimes in which the gauge and scalar fields are coupled via the electric permittivity. We find the conditions that allow for the presence of minimum energy configurations. In this formalism, the charge density must be written exclusively in terms of the components of the metric tensor and the scalar field is governed by first-order equations. We also find a manner to map the aforementioned equation into the corresponding one associated to kinks in $(1,1)$ spacetime dimensions, so we get analytical solutions for three specific spacetimes. We then calculate the energy density and show that the energy is finite. The stability of the solutions against contractions and dilations, following Derrick's argument, and around small fluctuations in the fields is also investigated. In this direction, we show that the solutions obeying the first-order framework are stable.
\end{abstract} 

\maketitle

\section{Introduction}
Scalar field models are of interest in several branches of Physics, such as in High Energy Physics \cite{weinberg}, Condensed Matter \cite{fradkin} and Cosmology \cite{bazeiacosm,defelice}. In particular, they can be used in the study of localized structures, such as kinks, vortices and monopoles \cite{manton}. Kinks are the simplest ones and arise under the canonical action of a single real scalar field in $(1,1)$ flat spacetime dimensions \cite{vachaspati}. These objects are stable around small fluctuations and minimize the energy of the system in the lines of the so-called BPS procedure \cite{b,ps}.

The study of localized structures in canonical models with a single real scalar field only leads to stable configurations in $(1,1)$ flat spacetime dimensions due to the Derrick-Hobart argument \cite{derrick,hobart}. There are some ways to evade this restriction. For instance, one can consider a potential with explicit dependence on the radial coordinate \cite{prl}, non-canonical models Refs.~\cite{kglobal} or complex scalar fields coupled to gauge fields \cite{no,thooft,polyakov}. In particular, the procedure introduced in Ref.~\cite{prl} allowed for the presence of stable (or at least metastable) radially symmetric solutions in $(d,1)$ flat spacetimes.

The idea in \cite{prl} was further extended in Ref.~\cite{morris1}, where a class of noncanonical potentials was introduced so one may obtain radially symmetric solutions of the BPS equations on distinct geometric backgrounds, such as the Schwarzschild, cosmic string and wormhole spacetimes. The evasion of Derrick's theorem was widely considered in the literature to study scalar field models on curved spacetimes; see Refs.~\cite{carloni,herdeiro1,delgrosso1,delgrosso2,danilo1,danilo2,herdeiro2,app1}. In particular, it was also used to establish the non-existence of self-gravitating solitons in Maxwell-scalar models in curved spacetimes \cite{herdeiro1} and to investigate scalar field solutions on Lifshitz spacetimes \cite{danilo1,danilo2} and hyperscaling violating geometries \cite{app1}.

Localized structures can also be investigated within the so-called Maxwell-scalar models. In this context, the coupling between the scalar and gauge fields usually occurs via a generalized permittivity. In Refs.~\cite{electricloc}, it was shown that electrically charged localized structures can be obtained with this framework for a single point charge; the solutions satisfy the Bogomol'nyi \cite{b} bound, so the energy of the system is minimized. This result was extended in Ref.~\cite{bazeiadipole} to the presence of dipoles in two spatial dimensions, whose stable localized configurations are attainable with bipolar coordinates. The study of continuous charged distributions was shown in \cite{eletro1d} to be more intricate, as there is no BPS formalism to give rise to first-order equations. In this direction, it was shown that the Maxwell-scalar model in a single spatial dimension can be effectively treated as a single scalar field model in the presence of impurities, where the impurity can be related to the charge density. Maxwell-scalar models may also be of interest in the study of scalarization \cite{herdeiroflat}, whose setup may serve as a toy model for the aforementioned phenomenon of charged black holes in generalized scalar-tensor models. In curved spacetimes, the evasion of Derrick's scaling argument may also be circumvented in Maxwell-scalar models \cite{PRLHer,herdeirocurved}. In Ref.~\cite{morris2}, models of the aforementioned type were investigated on spherically symmetric spacetimes, such as the Reissner-Nordstr\"om one.

Another possibility of current interest concerns the study of the strong interaction between quarks and gluons via holographic Einstein-Maxwell-scalar models. It is based on the gauge/gravity duality, which has several interesting uses \cite{GGD}, a specific one being the mapping of the QCD phase diagram at finite temperature onto a dual theory that describes charged and asymptotically anti-de Sitter black holes  in five dimensions. This subject has been recently reviewed in \cite{Noronha} and may stimulate new investigations, related to the present study. We can also add charged bosonic and fermionic fields to explore the study of boson \cite{PRLbo,LRR} and fermion \cite{PRDfe} stars. In the second case, one can consider a real or neutral scalar coupled to a charged fermion via standard Yukawa coupling, capable of circumventing no-go theorems that prevent the existence of solitonic solutions. In this work, however, we shall deal with Maxwell-scalar models immersed in radially-symmetric spacetimes, focusing on the construction of a first-order framework which allows for the existence of analytical configurations.  

We organize the investigation as follows. In Sec.~\ref{secthemodel} we study the equations of motions and the energy-momentum tensor. Considering only the presence of fixed charge distributions and the absence of currents, we develop a first-order framework based on the minimization of the energy that imposes the charge density to be related to the some of the components of the metric tensor. We verify the stability against contractions and dilations, in the lines of Derrick-Hobart argument. Also, we analyze how the solution behaves in the presence of small fluctuations in the scalar field. To show that our method is robust, in Sec.~\ref{secmodels}, we consider Tolman's VI \cite{tolman}, exponential \cite{expmetric1} and hyperscaling violating \cite{hyp1,hyp2,hyp3,hyp4,hyp5} geometries. We conclude the investigation in Sec.~\ref{secconclusions}, in which we make some final comments and discuss perspectives for future research.

\section{Maxwell-scalar models}\label{secthemodel}
We investigate the action that couples a gauge field $A_\mu$ with a real scalar field $\phi$ through the electric permittivity of the system in $(d,1)$ curved spacetime dimensions, with
\be\label{LL}
\Sc = \!\int\! d^Dx\sqrt{|g|}\left(-\frac{\epsilon(\phi)}{4}F_{\mu\nu}F^{\mu\nu} +\frac12\n_\mu\phi\n^\mu\phi - A_\mu j^\mu\right)\!,
\ee
where $D=d+1$. Here, $g$ is the determinant of the metric tensor $g_{\mu\nu}$, $\epsilon(\phi)$ denotes the electric permittivity, $F_{\mu\nu} = \n_\mu A_\nu - \n_\nu A_\mu$ represents the electromagnetic strength tensor, and $j^\mu$ is a current generated by external sources. By varying the action with respect to the scalar and gauge fields, we get the following equations of motion
\bes \label{eom}
\bal \label{scalareom}
\Box\phi +\frac14\frac{d\epsilon}{d\phi}F_{\mu\nu}F^{\mu\nu} &= 0,\\ \label{maxwelleom}
\n_\mu\left(\epsilon(\phi)F^{\mu\nu}\right) &= j^\nu,
\eal
\ees
where $\Box=g^{\mu\nu}\n_\mu\n_\nu$ is the d'Alambertian operator. We are interested in spacetimes with radial symmetry, so we take the line element
\be\label{ds2}
ds^2 = f(r)dt^2 -h(r)dr^2 -k(r)\omega_{ij}d\theta^id\theta^j,
\ee
where $\theta_i$, $i=1\ldots d-1$, represents the non-radial coordinates and the functions $f(r)$, $h(r)$ and $k(r)$ are non-negative functions. In addition, the determinant of the metric is $\sqrt{|g|}=\sqrt{\omega(\theta)f(r)h(r)k(r)^{d-1}}=\sqrt{\omega(\theta)\tau(r)}$, in which we have written the determinant of $\omega_{ij}$ as $\omega(\theta)$ and we have taken $\tau(r)=f(r)h(r)k(r)^{d-1}$. Considering static radially-symmetric configurations and the absence of currents, i.e., $j^0=\varrho(r)$ and $j^i=0$, we can define the electric charge associated to the system as $Q = \int_\Sigma d^dx\sqrt{\gamma}n_0j^0$, where $\gamma=|g|/f$ is the determinant of the induced metric on the surface $\Sigma$ defined at fixed $t$. Also, $n_0=\sqrt{f}$ is the temporal component of a vector $n^\mu$ normal to $\Sigma$. By expanding the aforementioned expression for the charge, one can show that
\be\label{charge}
Q = \Omega(d)\int_0^{\infty} dr\sqrt{\tau}\varrho(r),
\ee
with $\Omega(d) = \int_\Sigma d\theta_1\ldots d\theta_{d-1}\, \omega(\theta)$. From Eq.~\eqref{maxwelleom}, we get
\be
\begin{aligned}
\frac{1}{\sqrt{\tau}}{\big(\sqrt{\tau}\epsilon F^{r0}\big)}^\prime = \varrho(r),
\end{aligned}
\ee
in which the prime represents the derivative with respect to $r$. The above equation corresponds to the Gauss' law for our system; it can be solved by taking the charge density in the form
\be\label{densQ}
\varrho(r) =  \frac{\Qc^\prime(r)}{\sqrt{\tau(r)}},
\ee
where $\Qc(r)$ is an arbitrary function that obeys $\Qc(0)=0$ to avoid the presence of a Dirac delta in the charge density. In this case, the solution is
\be\label{eqa0}
F^{r0} = \frac{\Qc(r)}{\sqrt{\tau(r)}\epsilon(\phi)},
\ee
from which we get the gauge field $A^i=0$ and $\partial^r A^0=F^{r0}$. The presence of $\Qc(r)$ in the charge density \eqref{densQ} is interesting because it allows us to calculate the charge \eqref{charge} in a simple manner, as
\be\label{totalQ}
Q = \Omega(d)\Qc(\infty).
\ee
The intensity of the electric field can then be calculated by $|{\bf E}| = \sqrt{-F_{r0}F^{r0}}$, or
\be\label{Efield}
|{\bf E}| = \left|\frac{\Qc(r)}{\sqrt{k(r)^{d-1}}\,\epsilon(\phi)}\right|.
\ee
We use this in the equation of motion \eqref{scalareom}, that governs the scalar field, to get
\be\label{phiE}
\Box\phi +\frac{\Qc(r)^2}{k(r)^{d-1}}\frac{d}{d\phi}\left(\frac{1}{2\epsilon(\phi)}\right) = 0.
\ee
The energy-momentum tensor $\tensor{T}{^\mu_\nu}$ for non-static configurations is not conserved. Instead, it obeys $\nabla_\mu \tensor{T}{^\mu_\nu} =  j_\alpha \tensor{F}{^\alpha_\nu}$,
where
\be
\begin{aligned}
\tensor{T}{^\mu_\nu} &= \epsilon(\phi)\left(F^{\mu\lambda}F_{\lambda\nu} +\frac14\delta^\mu_\nu F_{\lambda\sigma}F^{\lambda\sigma}\right) \\
& +\n^\mu\phi\n_\nu\phi -\frac12\delta^\mu_\nu\n_\lambda\phi\n^\lambda\phi.
\end{aligned}
\ee
In spite of this, if the fields are static, the energy density is conserved, i.e., $\nabla_0 \tensor{T}{^0_0}=0$. Considering that $\phi$ depends only on $r$ with $j^0=\varrho(r)$ and $j^i=0$, the only non-null components of the energy-momentum tensor are the energy density
\be\label{t00geral}
\begin{aligned}
\tensor{T}{^0_0} &= \frac{\epsilon(\phi)}{2}\,|{\bf E}|^2  +\frac{1}{2h(r)}\,{\phi^{\prime}}^2\\
    &= \frac{\Qc(r)^2}{2k(r)^{d-1}\epsilon(\phi)} +\frac{1}{2h(r)}\,{\phi^{\prime}}^2,
\end{aligned}
\ee
and the stresses. The radial stress is
\bes\be\label{stressr}
\begin{aligned}
\tensor{T}{^r_r} &= \frac{\epsilon(\phi)}{2}|{\bf E}|^2 -\frac{1}{2h(r)}{\phi^{\prime}}^2\\
    &= \frac{\Qc(r)^2}{2k(r)^{d-1}\epsilon(\phi)} -\frac{1}{2h(r)}{\phi^{\prime}}^2,
\end{aligned}
\ee
and the stress in the other directions $\theta_i$ defined in Eq.~\eqref{ds2} is
related to the radial one, in the form
\be\label{stresstheta}
\tensor{T}{^{\theta_i}_{\theta_j}}=-\delta^{i}_{j}\tensor{T}{^r_r}.
\ee
\ees
Static fields and charges and the absence of currents make Eq.~\eqref{phiE} become
\be\label{eomphistatic}
-\frac{1}{\sqrt{\tau(r)}}\left(\frac{\sqrt{\tau(r)}}{h(r)}\phi^\prime\right)^\prime +\frac{\Qc(r)^2}{k(r)^{d-1}}\frac{d}{d\phi}\left(\frac{1}{2\epsilon(\phi)}\right) = 0.
\ee
By integrating the energy density, we get the energy in the form
\be\label{energy}
\begin{aligned}
\Ec &= \int_0^\infty d^dx\sqrt{|g|}\tensor{T}{^0_0}\\
    &= \Omega(d)\!\int_0^\infty\!\!dr\sqrt{\tau(r)}\left(\frac{\Qc(r)^2}{2k(r)^{d-1}\epsilon(\phi)} +\frac{1}{2h(r)}{\phi^{\prime}}^2\right)\!.
\end{aligned}
\ee
This expression can be rearranged in the form
\be
\begin{aligned}
\Ec \!&=\! \Omega(d)\!\int_0^\infty\!\! dr\!\left(\frac{\sqrt{\tau}}{2h}\!\left(\!\phi^{\prime} \mp\sqrt{\frac{h\Qc^2}{k^{d-1}\epsilon(\phi)}}\right)^{\!2}\! \pm\sqrt{\frac{f\Qc^2}{\epsilon(\phi)}}\phi^{\prime}\!\right)\!.
\end{aligned}
\ee
In order to write the term inside the integral as a total derivative, we follow the lines of Refs.~\cite{b,ps} and introduce the auxiliary function $W=W(\phi)$ such that
\be\label{condbps}
\Qc(r)=\frac{e}{\sqrt{f(r)}}\quad\text{and}\quad
\frac{e}{\sqrt{\epsilon(\phi)}} = \frac{dW}{d\phi},
\ee
with $e$ being a constant that we consider to be positive without loss of generality. The expression on the left shows that the charge is now constrained to the metric. Since we have imposed $\Qc(0)=0$, we see that the factor of $dt^2$ in the line element \eqref{ds2} must obey $f(0)\to\infty$. 
Note that this condition precludes the use of backgrounds where the function $f(r)$ is smooth at $r=0$, such as de Sitter and others, although $f(r)$ could tend to de Sitter in domains that do not include the small $r$ region. 
 Using both equations in Eq.~\eqref{condbps}, we can rewrite the energy as
\be
\begin{aligned}
\Ec &= \Omega(d)\int_0^\infty dr\left(\frac{\sqrt{\tau}}{2h}\left(\phi^{\prime} \mp\frac{h}{\sqrt{\tau}}\frac{dW}{d\phi}\right)^2\right) + \Ec_B,
\end{aligned}
\ee
where $\Ec_B = \Omega(d)|W(\phi(\infty))-W(\phi(0))|$. The above expression allows us to see that the energy has a lower bound, $\Ec\geq \Ec_B$. Therefore, the solutions which minimize the energy of the system obey the first-order equations
\be\label{fo}
\phi^{\prime} = \pm\frac{h}{\sqrt{\tau}}\frac{dW}{d\phi},
\ee
which are compatible with the equation of motion \eqref{eomphistatic}. It is worth highlighting that the solutions of the above first order equations are stressless, that is, $\tensor{T}{^r_r}=\tensor{T}{^{\theta_i}_{\theta_j}}=0$; see Eqs.~\eqref{stressr} and \eqref{stresstheta}.

To search for analytical solutions, we can define the variable $x$ as
\be\label{dx}
dx = \frac{h(r)}{\sqrt{\tau(r)}}dr = \sqrt{\frac{h(r)}{f(r)k^{d-1}(r)}}\,dr
\ee
to write the first-order equations \eqref{fo} in the form
\be\label{fox}
\frac{d\phi}{dx} = \pm\frac{dW}{d\phi}.
\ee
The above framework is the same that arises with the standard Lagrangian density of a single scalar field, $\LL=\frac12\partial_\mu\phi\partial^\mu\phi - V(\phi)$, in $(1,1)$ spacetime dimensions, with the scalar potential being $V(\phi)=\frac12(dW/d\phi)^2$. Therefore, we are now able to map kink-like one-dimensional solutions to our model in the curved spacetime. This mapping, however, requires the need to be careful with the choice of the functions $f(r)$, $h(r)$ and $k(r)$, as the solution must be fully mapped into $r$ as it varies from $0$ to $\infty$. Moreover, although there is a correspondence between scalar field solutions of the model \eqref{LL} and the corresponding one in $(1,1)$ spacetime dimensions, the current model engenders electric field with intensity \eqref{Efield}. In Ref.~\cite{electricloc}, for instance, one provides Maxwell-scalar systems of a single point charge in flat spacetime that support first-order equations which lead to stable finite energy solutions.

With the condition in the left equation of \eqref{condbps}, we can rewrite the charge density \eqref{densQ} in the form 
\be\label{densQbps}
\varrho(r) =  -\frac{e f'(r)}{2f^2(r)\sqrt{h(r)k^{d-1}(r)}}.
\ee
The charge in Eq.~\eqref{totalQ} reads
\be\label{Qbps}
Q = \frac{e\,\Omega(d)}{\sqrt{f(\infty)}}.
\ee
This expression shows that the charge of the system depends exclusively on the behavior of the $f(r)$ at infinity. For $f(\infty)\to0$, the charge diverges; for $f(\infty)\to f_\infty$, with $f_\infty$ being a non-null constant, the charge is finite and non zero; for $f(\infty)\to \infty$ the charge vanishes. Taking Eqs.~\eqref{condbps} into account, the energy density in Eq.~\eqref{t00geral} becomes
\be\label{t00bps}
\tensor{T}{^0_0}=\frac{1}{f(r)k(r)^{d-1}}\left(\frac{dW}{d\phi}\right)^2.
\ee
The contribution of the scalar field does not lead to divergences in the above expression since we are considering the change of variables \eqref{dx} and imposing that the solutions connect the minima of $V(\phi)=\frac12(dW/d\phi)^2$ as $r$ varies from $0$ to $\infty$. Therefore, divergences in the above equation may only arise due to the metric.

\subsection{Stability}
We now investigate the stability of our model. First, we study how the energy of the system behaves by rescaling the solution of the equation of motion \eqref{eomphistatic}, following the lines of Derrick-Hobart argument \cite{derrick,hobart}. By taking $r\to r^{(\lambda)}=\lambda r$ in the scalar field solution, we have $\phi(r)\to\phi^\lambda(r)=\phi(\lambda r)$. We denote energy of the rescaled solution, which we calculate from Eq.~\eqref{energy} as $\Ec^{(\lambda)}$. The solution is stable against contractions and dilations if $\lambda=1$ is the minimum of $\Ec^{(\lambda)}$. By taking $\partial\Ec^{(\lambda)}/\partial\lambda|_{\lambda=1}=0$ and $\partial^2\Ec^{(\lambda)}/\partial\lambda^2|_{\lambda=1}>0$, we get the conditions
\bes\label{derrick}\be\label{derrick1}
\int_0^\infty\!  dr\Bigg[\left(\frac{r\sqrt{\tau}\Qc^2}{k^{d-1}}\right)^{\!\prime}\frac{1}{\epsilon(\phi)} +r^2\left(\frac{\sqrt{\tau}}{rh}\right)^{\!\prime}{\phi^{\prime}}^2\Bigg] = 0,
\ee
\be\label{derrick2}
\begin{aligned}
\int_0^\infty\!  dr\Bigg[\left(\frac{r^2\sqrt{\tau}\Qc^2}{k^{d-1}}\right)^{\!\prime\prime}\frac{1}{\epsilon(\phi)} +r^2\left(\frac{\sqrt{\tau}}{h}\right)^{\!\prime\prime}{\phi^{\prime}}^2\Bigg] > 0.
\end{aligned}
\ee
\ees
The integral in Eq.~\eqref{derrick1} must be zero, so it is a \emph{global} condition. Since the function inside the integral may change its sign and depends explicitly on $r$, we cannot take it to be zero locally. Notwithstanding that, an interesting issue occurs: by considering the conditions \eqref{condbps} and the first-order equation \eqref{fo}, one can show that the integrand vanishes \emph{locally} and the condition \eqref{derrick1} becomes an identity. Moreover, in this situation, we can write Eq.~\eqref{derrick2} in the form
\be
\int^\infty_0 dr\Bigg[\frac{\sqrt{\tau}}{h}{\left(\frac{yh}{\sqrt{\tau}}\right)^{\!\prime}}^2\left(\frac{dW}{d\phi}\right)^2\Bigg] > 0,
\ee
which is always satisfied, since the function inside the integral is non negative. Hence, the solutions of the first-order equations are stable against contractions and dilations. Moreover, these solutions also lead to the absence of stress as one can verify from Eqs.~\eqref{stressr} and \eqref{stresstheta}.

Next, let us verify the stability of the solutions $A^0(r)$ and $\phi(r)$ of the equations of motion \eqref{eqa0} and \eqref{eomphistatic} around small fluctuations. By taking $\phi(x^\alpha)=\phi(r)+\eta(x^\alpha)$ and $A_\mu(x^\alpha)= A_\mu(r)+\Ac_\mu(x^\alpha)$ we can write $F_{\mu\nu}(x^\alpha)= F_{\mu\nu}(r)+\Fc_{\mu\nu}(x^\alpha)$, where $\Fc_{\mu\nu}=\partial_\mu\Ac_\nu-\partial_\nu\Ac_\mu$. By substituting this into the equation of motion \eqref{eom}, we get
\bes\bal\label{flucscalar}
&\Box\eta +\frac14\frac{d^2\epsilon}{d\phi^2}F_{\mu\nu}F^{\mu\nu}\eta +\frac12\frac{d\epsilon}{d\phi}F_{\mu\nu}\Fc^{\mu\nu} = 0,\\
&\n_\mu\left(\frac{d\epsilon}{d\phi}F^{\mu\nu}\eta +\epsilon(\phi)\Fc^{\mu\nu}\right) = 0.
\eal
\ees
The latter equation admits the solution
\be\label{Fpertub}
\Fc^{\mu\nu} = -\frac{d\ln(\epsilon)}{d\phi}F^{\mu\nu}\eta,
\ee
showing that the fluctuations of the gauge and scalar fields are related. By substituting the above expression into Eq.~\eqref{flucscalar}, we get
\be
\Box\eta(x^\alpha) +\frac{\Qc(r)^2}{k(r)^{d-1}}\frac{d^2}{d\phi^2}\left(\frac{1}{2\epsilon}\right)\eta(x^\alpha) = 0.
\ee
We then take the fluctuations in the form $\eta(x^\alpha) =\sum_i\eta_i(r)\cos(\omega_it)$, such that the above equation becomes
\be\label{stabeq}
-\frac{1}{\sqrt{\tau}}\left(\frac{\sqrt{\tau}}{h}\eta_i^\prime\right)^\prime +\frac{\Qc(r)^2}{k(r)^{d-1}}\frac{d^2}{d\phi^2}\left(\frac{1}{2\epsilon}\right)\eta_i = \frac{\omega_i^2}{f}\eta_i.
\ee
This expression is an eigenvalue equation of the Sturm-Liouville type. The solutions are stable if the eigenvalues are non negative, i.e., $\omega_i^2\geq0$. If the static solution $\phi(r)$ obeys the first-order framework \eqref{condbps}--\eqref{fo}, we can write the above equation as: $-\left(\sigma(r)s(r)^2\eta_i^\prime\right)^\prime+\sigma(r) U(r)\eta_i = \omega_i^2\sigma(r)\eta_i$, or $L\eta_i = \omega^2_i\eta_i$, where $L$ is the Sturm-Liouville operator given by
\be\label{stab}
L = -\frac{1}{\sigma(r)}\frac{d}{dr}\sigma(r)s(r)^2\frac{d}{dr} +U(r),
\ee
with
\bes\label{sigmasu}
\bal\label{sigmasl}
\sigma(r) &= \frac{h(r)k(r)^{d-1}}{\sqrt{\tau(r)}} = \sqrt{\frac{h(r)k(r)^{d-1}}{f(r)}},\\ \label{ssl}
s(r) &= \frac{\sqrt{\tau(r)}}{h(r)\sqrt{k(r)^{d-1}}} = \sqrt{\frac{f(r)}{h(r)}},\\ \label{usl}
U(r) &= \frac{1}{k(r)^{d-1}}\left(\left(\frac{d^2W}{d\phi^2}\right)^2 +\frac{dW}{d\phi}\frac{d^3W}{d\phi^3}\right).
\eal
\ees
In this specific situation, we were able to show that eigenvalue equation associated to the operator \eqref{stab} supports a zero mode ($\omega_0=0$) related to the derivative of the static solution,
\be\label{zeromode}
\eta_0(r) = \Nc\frac{\sqrt{\tau}}{h}\phi^\prime,
\ee
in which $\Nc$ is a constant of normalization that is obtained from
\be\label{condnorm}
\int^\infty_0 dr\,\sigma(r)\eta_0^2(r) = 1.
\ee
If there are no negative modes, the fluctuations do not destabilize the scalar solutions. We then investigate how the gauge field behaves under these fluctuations by using Eq.~\eqref{Fpertub} to calculate the term $F_{\mu\nu}(x^\alpha)F^{\mu\nu}(x^\alpha)$ and get
\be
\begin{aligned}
    -F_{\mu\nu}(x^\alpha)F^{\mu\nu}(x^\alpha) &= \frac{2}{fk^{d-1}}\!\left(\frac{dW}{d\phi}\right)^2\!\!\left(\!1 -\frac{d\ln(\epsilon)}{d\phi}\,\eta(x^\alpha)\!\right)\!,
\end{aligned}
\ee
where the scalar fluctuations are $\eta(x^\alpha) =\sum_i\eta_i(r)\cos(\omega_it)$, as considered right above Eq.~\eqref{stabeq}. We then see that the functions which control the metric and the scalar field must be chosen to avoid divergences in the above expression to keep the gauge field stable.

The Sturm-Liouville operator \eqref{stab} can be factorized in terms of adjoint operators $S$ and $S^\dagger$, as $L=S^\dagger S$, where
\bes\label{selfadjoint}
\bal
S &= s(r)\!\left(-\frac{d}{dr} +\frac{h(r)}{\sqrt{\tau(r)}\phi^\prime}\left(\frac{\sqrt{\tau(r)}\phi^\prime}{h(r)}\right)^\prime\right),\\
S^\dagger &= s(r)\!\left(\frac{d}{dr} +\frac{h(r)}{\sqrt{\tau(r)}\phi^\prime}\left(\frac{\sqrt{\tau(r)}\phi^\prime}{h(r)}\right)^\prime +\frac{\big(\sigma(r)s(r)\big)^\prime}{\sigma(r)s(r)}\right)\!.
\eal
\ees
The above factorization and the absence of nodes in the zero mode ensures that the Sturm-Liouville operator only admits non-negative eigenvalues, so the solutions are stable against small fluctuations. To get another interpretation of the stability, we can transform the stability equation into a Schr\"odinger-like one with the change of variables
\be\label{change}
dy = \frac{dr}{s} \quad\text{and}\quad \psi_i = \sqrt{\sigma s}\,\eta_i.
\ee
This makes the eigenvalue equation have the form $\mathcal{H}\eta_i = \omega_i^2\eta_i$, where
\be\label{hu}
\mathcal{H} = -\frac{d^2}{dy^2} + \Uc(y),\quad\text{with}\quad
\Uc(y) = \frac{(\sqrt{\sigma s})_{yy}}{\sqrt{\sigma s}} +U(r(y)),
\ee
in which $r(y)$ is the coordinate $r$ written in terms of the variable $y$ defined in Eq.~\eqref{change}. The zero mode of the Schr\"odinger-like equation \eqref{hu} is then given by
\be
\psi_0(y) = \Nc k(r(y))^{\frac34(d-1)}\phi_y.
\ee
It is worth commenting that, even though we are able to get the description of the linear stability using a Schr\"odinger-like equation, obtaining the analytical expressions for the change of variables \eqref{change} is not always possible. However, one can use numerical methods to circumvent this issue when necessary. 

Next, we investigate specific models with the metric \eqref{ds2} that support stable electrically charged localized structures.

\section{Models}\label{secmodels}
To illustrate our procedure, we must consider scalar field models that allow for the presence of localized solutions which connect the minima of $V(\phi)=\frac12(dW/d\phi)^2$. Due to the nature of the geometric backgrounds that we shall investigate, we consider the model introduced in Ref.~\cite{prl}, described by the auxiliary function
\be\label{wpmodel}
W(\phi) = \frac{p}{2p-1}\phi^{2-\frac1p} -\frac{p}{2p+1}\phi^{2+\frac1p},
\ee
with $p=3,5,7,\ldots$. The permittivity can be calculated from the right equation in \eqref{condbps}, which leads us to $\epsilon(\phi) = e^2\phi^{-2}(\phi^{-1/p}-\phi^{1/p})^{-2}$. This function diverges at $\phi=\pm1$ and $\phi=0$. It has two sectors, $\phi\in[-1,0]$ and $\phi\in[0,1]$.

Supposing that $\Qc$ is given by the left equation in \eqref{condbps}, we can use the first-order framework, such that the scalar field is governed by Eq.~\eqref{fo}, which reads
\be
\frac{d\phi}{dx} = \pm\phi\left(\phi^{-1/p}-\phi^{1/p}\right).
\ee
This equation admits the solution
\be\label{solp2kink}
\phi_\pm(x) = \pm\tanh^p\!\left(x/p\right),
\ee
where $x$ must be calculated from \eqref{dx} for the specific metric under investigation. The upper/lower sign represents the increasing/decreasing solution. This solution engenders an inflection point with null derivative at $x=0$ and is extended, attaining the boundary values only at $\pm\infty$. It goes from $\phi=-1$ to $\phi=1$, passing through $\phi=0$, which separates the sectors, due to the atypical character of this point in the scalar potential $V(\phi)=\frac12(dW/d\phi)^2=\frac12\phi^2(\phi^{-1/p}-\phi^{1/p})^2$ associated to the model in $(1,1)$ dimensions to be mapped (see Ref.~\cite{prl}). The presence of the inflection point with null derivative in $\phi_\pm(x)$ allows us to split it into the half-compact solutions
\bes\label{solp}
\be\label{solp1}
\phi_\pm(x)=\begin{cases}
    \pm\tanh^p\!\left( \cfrac{x}{p}\right), & x\leq0\\
    0, & x>0
\end{cases}
\ee
and
\be\label{solp2}
\phi_\pm(x)=\begin{cases}
    0,& x<0\\
    \pm\tanh^p\!\left( \cfrac{x}{p}\right), & x\geq0.
\end{cases}
\ee
\ees
These solutions will be used in situations where the geometric background induces a mapping of $r$ into $x\in(-\infty,0)$ and $x\in(0,\infty)$. A similar technique was already applied with this model in Refs.~\cite{app1,app2}. We remark that, only if the change of variables in Eq.~\eqref{dx} leads to $x\in(-\infty,\infty)$, the case $p=1$ is allowed, as the solution will connect the minima $\phi=-1$ and $\phi=1$ of the scalar potential.

In the following subsections, we apply our method in three illustrative geometries. 

\subsection{Tolman's metric}
First, we investigate the behavior of our model on the Tolman's metric VI, introduced in Ref.~\cite{tolman}, with the line element
\be\label{ds2tolman}
\begin{aligned}
    ds^2 &= (Ar^{1-n} + Br^{1+n})^2 dt^2 - (2-n^2)dr^2\\
    &-r^2\,d\theta^2 -r^2\sin^2\theta\,d\vphi^2,
\end{aligned}
\ee
where $A$ and $B$ are positive real parameters, and $n$ is a real number that obeys $|n|<\sqrt{2}$. This metric was originally derived to obtain exact perfect fluid solutions, and represents a sphere with infinite pressure and density at the center. One thus must assume that the constants $A$ and $B$ are positive and that no event horizon exists. Since radial geodesics reach the center in finite affine time and curvature scalars diverge there, in modern terms we would say that this represents a naked singularity. It should be evident that the line element (\ref{ds2tolman}) is not asymptotically flat, as it is expected to describe the interior of stellar objects. A realistic spacetime should consider an exterior line element (such as Schwarzschild, Schwarzschild-de Sitter, or others) that matches this one at some surface $r=R_0$. Since we are only interested in the illustration of the applicability of our formalism, we will not consider such refinements here, though they will be explored in a separate publication. 

\begin{figure}[t!]
    \centering
    \includegraphics[width=4.2cm]{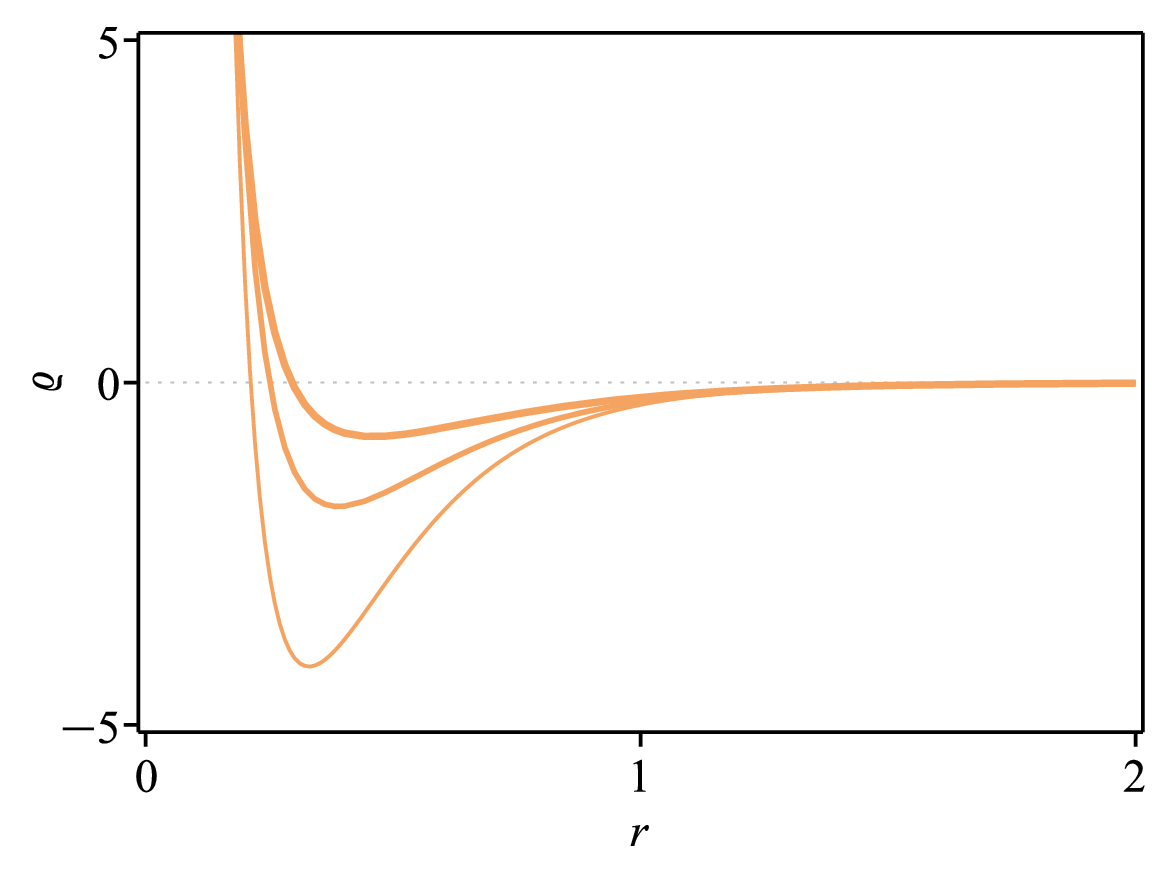}
    \includegraphics[width=4.2cm]{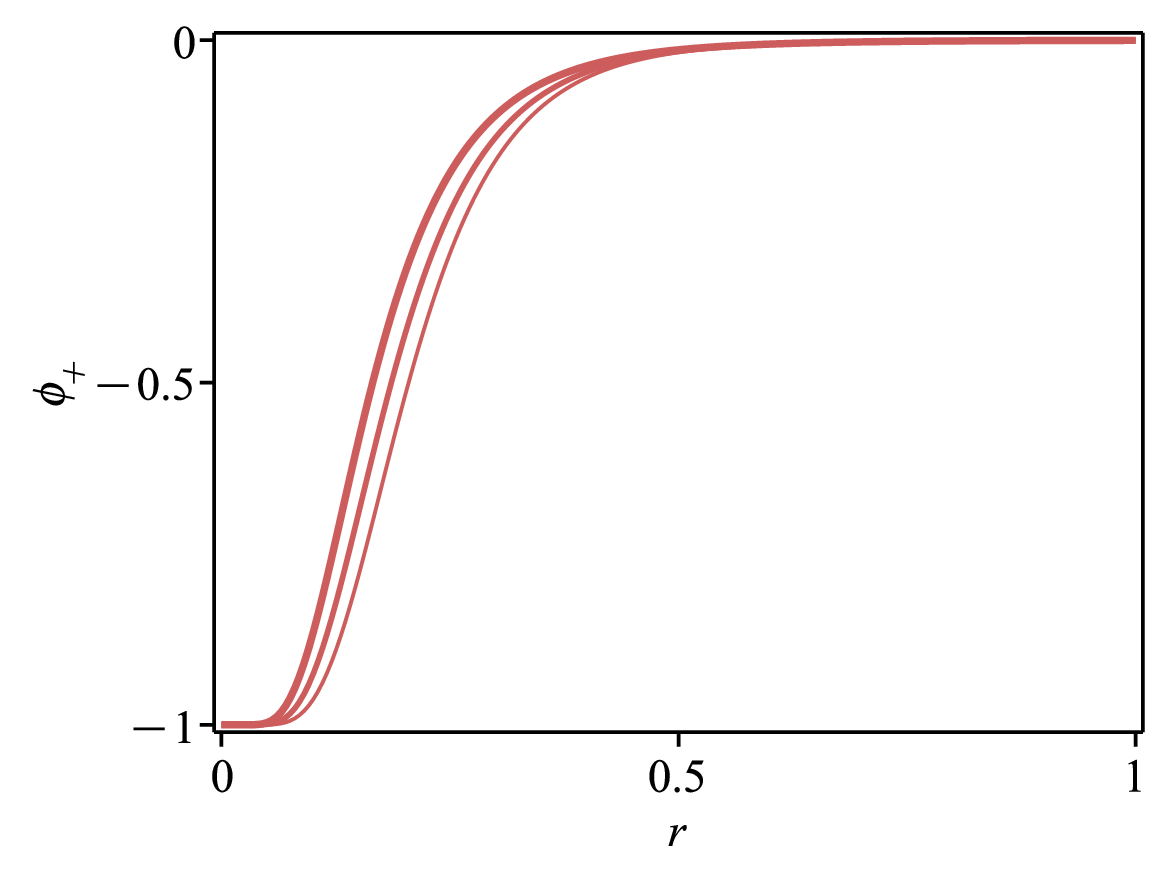}
    \includegraphics[width=4.2cm]{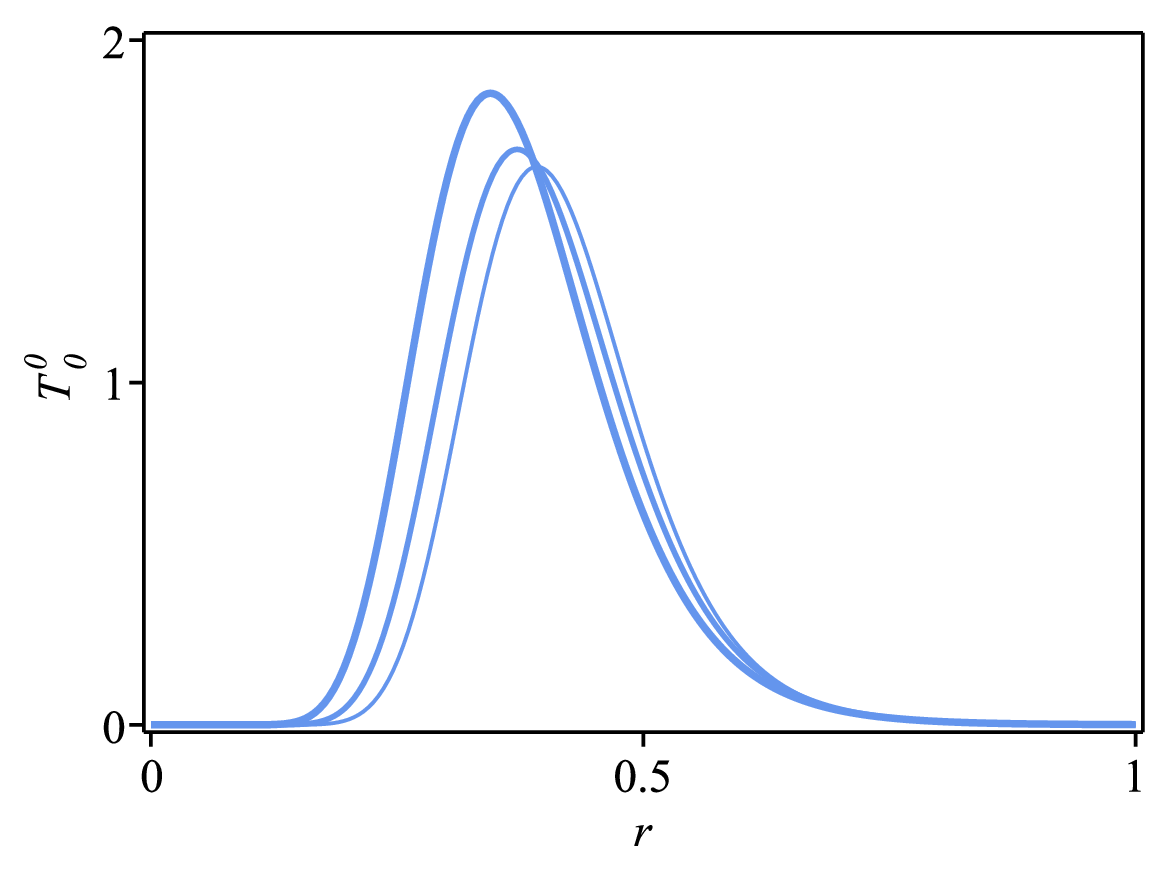}
    \includegraphics[width=4.2cm]{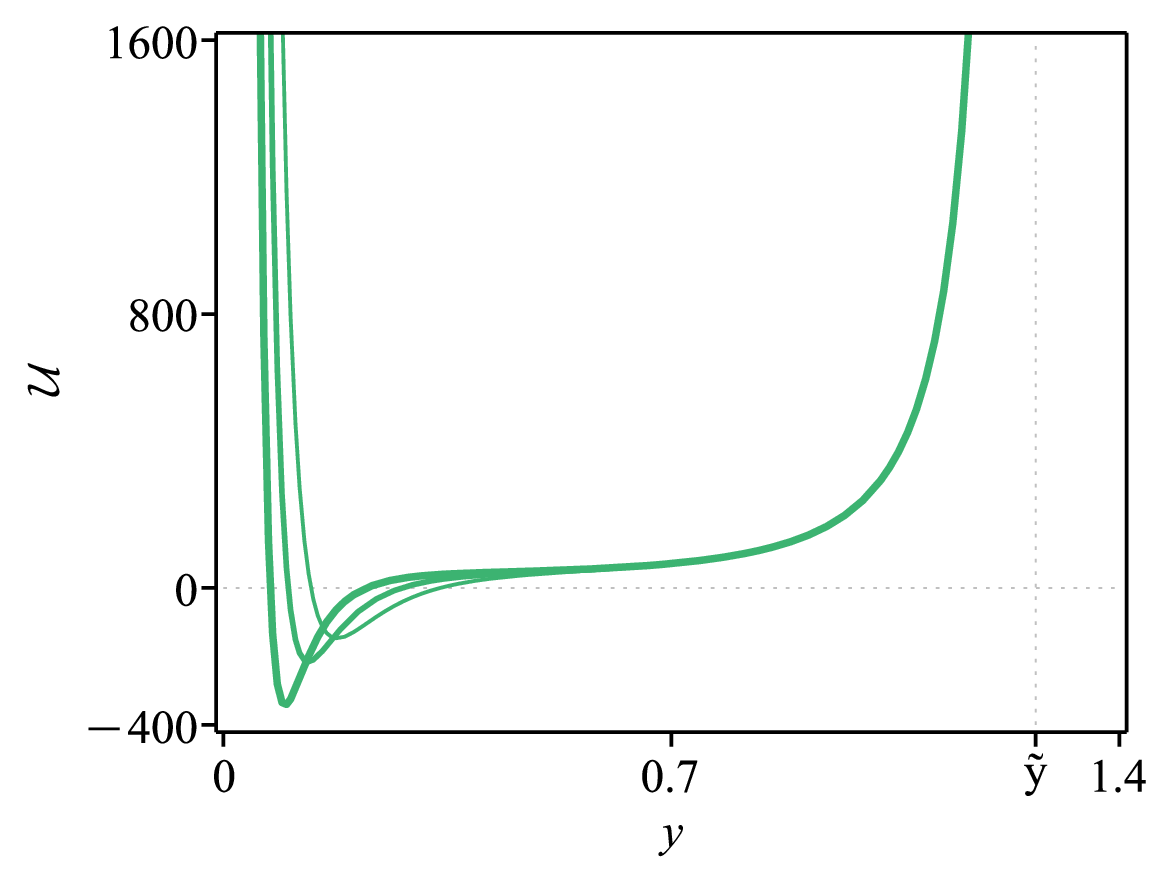}
    \caption{The charge density $\varrho(r)$ in Eq.~\eqref{rhotolman} (top left, orange), solution $\phi_+(r)$ in Eq.~\eqref{soltolman} (top right, red), energy density $\tensor{T}{^0_0}(r)$ in Eq.~\eqref{t00tolman} (bottom left, blue) and the Schr\"odinger-like potential $\Uc(y)$ in Eq.~\eqref{Uytolman} (bottom right, green), for $e=1$, $p=3$, $n=1.1$, $A=B^{-1}=5/6,1,6/5$. In each panel, the thickness of the lines increases with increasing $A$. }
    \label{fig1}
\end{figure}

The above expression (\ref{ds2tolman}) leads to $\sqrt{|g|} = \sqrt{2-n^2}\,r^2(Ar^{1-n} + Br^{1+n})\sin\theta$ and $\sqrt{\tau} = \sqrt{2-n^2}\,r^2(Ar^{1-n} + Br^{1+n})$. In the first-order framework, one requires that Eq.~\eqref{condbps} must be obeyed, from which we get
\be\label{qc1}
\Qc(r) = \frac{e}{Ar^{1-n} + Br^{1+n}}.
\ee
Since we must have $\Qc(0)=0$, the parameter $n$ is now restricted to $1<|n|<\sqrt{2}$. The charge density in Eq.~\eqref{densQbps} reads
\be\label{rhotolman}
\varrho(r)=-\frac{e\left(A(1-n)r^{-n} + B(1+n)r^{n}\right)}{\sqrt{2-n^2}\,r^2\left(Ar^{1-n} + Br^{1+n}\right)^3}.
\ee
It is zero for $r=r_*$, with $r_*=\left(A(n-1)/(B(n+1))\right)^{1/(2n)}$. Interestingly, $\varrho(r)$ is positive for $r<r_*$ and negative for $r>r_*$. Asymptotically, the charge density behaves as $\varrho(r) \propto r^{-2n-5}$, which always goes to zero for $1<|n|< \sqrt{2}$. Near the origin, we see that $\varrho(r)\propto r^{2n-5}$. Therefore the charge density has a divergence at $r=0$ for $n$ in the aforementioned interval. Nevertheless, this divergence is integrable, as the charge in Eq.~\eqref{Qbps} is null. We display the charge density in the top-left panel of Fig.~\ref{fig1}.

To obtain the scalar field solution, we use the change of variables in Eq.~\eqref{dx}, which leads us to
\be\label{xtolman}
\begin{aligned}
x(r) &= -\frac{\sqrt{2-n^2}}{A(2-n)}\!\bigg(\!r^{n-2}\,{}_2F_1\left(1,\frac12-\frac1n;\frac32-\frac1n;-\frac{B}{A}\,r^{2n}\right)\\
    &+\frac{\pi(2-n)}{2n}\left(\frac{B}{A}\right)^{(2-n)/n}\sec\left(\frac{\pi}{n}\right)\bigg),
\end{aligned}
\ee
where ${}_2F_1(a,b;c;z)$ represents the hypergeometric function of argument $z$. Notice that the above expression maps the interval $r\in(0,\infty)$ into $x\in(-\infty,0)$. Therefore, we must use the solution in Eq.~\eqref{solp1}, so we have
\be\label{soltolman}
\phi_\pm(r) = \pm\tanh^p\left(\frac{x(r)}{p}\right).
\ee
The upper/lower sign represents the increasing/decreasing solution and $x(r)$ is as in Eq.~\eqref{xtolman}. Notice that $\phi_+$ connects $\phi=-1$ at $r=0$ to $\phi=0$ at $r\to\infty$ and $\phi_-$ goes from $\phi=1$ at $r=0$ to $\phi=0$ at $r\to\infty$. To show the behavior of the above solution, we display $\phi_+(r)$ for some values of the parameters in the top-right panel of Fig.~\ref{fig1}. The energy density \eqref{t00bps} associated to both $\phi_+$ and $\phi_-$ is
\be\label{t00tolman}
\tensor{T}{^0_0}(r) = \frac{\sech^4(x(r)/p)\tanh^{2(p-1)}(x(r)/p)}{r^4\left(Ar^{1-n}+Br^{1+n}\right)^2},
\ee
from which we see that $\tensor{T}{^0_0}(0)=0$. It is plotted in the bottom-left panel of Fig.~\ref{fig1}. By integrating this expression, we get the energy $\Ec=2p\Omega(d)/(4p^2-1)$, as expected from the definition of $\Ec_B$ above Eq.~\eqref{fo}. Note from Fig.~\ref{fig1} that both the charge density and the energy density are localized within the innermost region of the spacetime, confirming in this way that we are dealing with localized structures. This indicates that there is no need to consider an explicit form for the exterior geometry and that matching to a Schwarzschild space-time or to other exterior line element would not significantly alter the solitonic character of these solutions.

The stability is governed by the Sturm-Liouville operator \eqref{stab}, which is described by the functions in Eqs.~\eqref{sigmasl} and \eqref{ssl} that lead us to
\be
\sigma(r) = \frac{\sqrt{2-n^2}\,r}{Ar^{-n}+Br^{n}}\quad\text{and} \quad s(r) = \frac{Ar^{1-n}+Br^{1+n}}{\sqrt{2-n^2}}.
\ee
From these expressions, we get that $\sigma(r)$ vanishes at the origin and diverges at infinity, and $s(r)$ diverges both for $r=0$ and $r\to\infty$. The stability potential \eqref{usl} reads
\be
\begin{aligned}
U(r) &= \frac{1}{r^4}\bigg(\left(1+\frac{1}{p}\right)\left(1+\frac{2}{p}\right)\tanh^2\left(\frac{x(r)}{p}\right)\\
&+\left(1-\frac{1}{p}\right)\left(1-\frac{2}{p}\right)\tanh^{-2}\left(\frac{x(r)}{p}\right)-2\bigg).
\end{aligned}
\ee
The zero mode associated to the stability equation can be calculated analytically from Eq.~\eqref{zeromode}; it is
\be
\eta_0(r) = \Nc\tanh^{p-1}\left(\cfrac{x(r)}{p}\right)\sech^2\left(\cfrac{x(r)}{p}\right),
\ee
where $\Nc$ is a normalization constant that cannot be obtained analytically for general values of the parameter $p$. Notwithstanding that, by using the asymptotic behavior of the above expression, we have verified that the zero mode can be normalized, so there is a finite value of $\Nc$ compatible with Eq.~\eqref{condnorm}. The absence of nodes in the zero mode shows that negative eigenvalues are absent, ensuring the stability of the model around small fluctuations.

Since we are following the first-order framework, the Sturm-Liouville operator can be factorized into the product of the adjoint operators \eqref{selfadjoint}, which take the form 
\bes
\bal
S &= -\frac{Ar^{1-n}+Br^{1+n}}{\sqrt{2-n^2}}\bigg(\frac{d}{dx} +\frac{\sqrt{2-n^2}}{r^2\left(Ar^{1-n}+Br^{1+n}\right)}\nn
    &\times\!\left(\!\left(\frac{1}{p}+1\!\right)\!\tanh\!\left(\frac{x(r)}{p}\right) \!+\!\left(\frac{1}{p}-1\!\right)\!\tanh^{-1}\!\left(\frac{x(r)}{p}\right)\!\right)\!\!\bigg),\\
S^\dagger &= \frac{Ar^{1-n}+Br^{1+n}}{\sqrt{2-n^2}}\bigg(\frac{d}{dx} -\frac{\sqrt{2-n^2}}{r^2\left(Ar^{1-n}+Br^{1+n}\right)}\nn
    &\times\!\left(\!\left(\frac{1}{p}+1\!\right)\!\tanh\!\left(\frac{x(r)}{p}\right) \!+\!\left(\frac{1}{p}-1\!\right)\!\tanh^{-1}\!\left(\frac{x(r)}{p}\right)\!\right)\nn
    &+\frac{2}{r}\bigg).
\eal
\ees
The stability equation \eqref{stabeq} can be transformed into a Schr\"odinger-like one with the change of variables \eqref{change},
\be\label{yr1}
y = \frac{\sqrt{2-n^2}}{n\sqrt{AB}}\arctan\left(\sqrt{\frac{B}{A}}\,r^n\right),
\ee
with $y\in\big[0,\tilde{y}\big]$, where $\tilde{y}=\pi\sqrt{2-n^2}/(2n\sqrt{AB})$. The potential associated to the operator \eqref{hu} can be written as
\be\label{Uytolman}
\begin{aligned}
\Uc(y) &= U(r(y)) +\frac{4AB}{2-n^2}\csc^2\left(\frac{2n\sqrt{AB}\,y}{\sqrt{2-n^2}}\right)\\
    &\times\left(1-n\cos\left(\frac{2n\sqrt{AB}\,y}{\sqrt{2-n^2}}\right)\right),
\end{aligned}
\ee
for $y\in\big[0,\tilde{y}\big]$. Notice that $\Uc(y)$ diverges for $y\to \tilde{y}$. In the above equation, $r(y)$ denotes the inverse of the expression in Eq.~\eqref{yr1}. This potential is an infinite well of width $\tilde{y}$, so it only admits bound states. In the bottom-right panel of Fig.~\ref{fig1} one can see the behavior of the above stability potential for some values of the parameters. Notice that there is a region where $\Uc(y)$ is negative, allowing for the presence of the zero mode. The zero mode is the ground state, so the solution is linearly stable.

\subsection{Exponential metric}
\begin{figure}[t!]
    \centering
    \includegraphics[width=4.2cm]{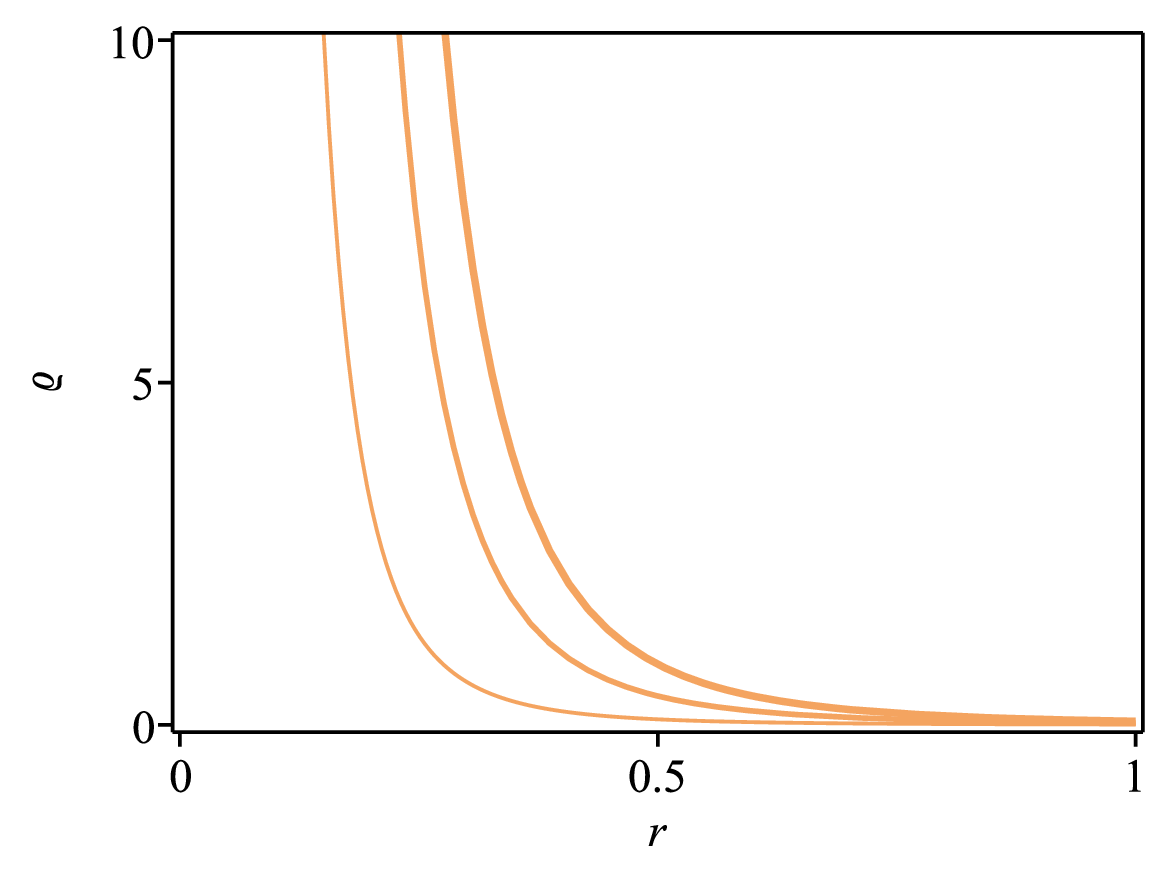}
    \includegraphics[width=4.2cm]{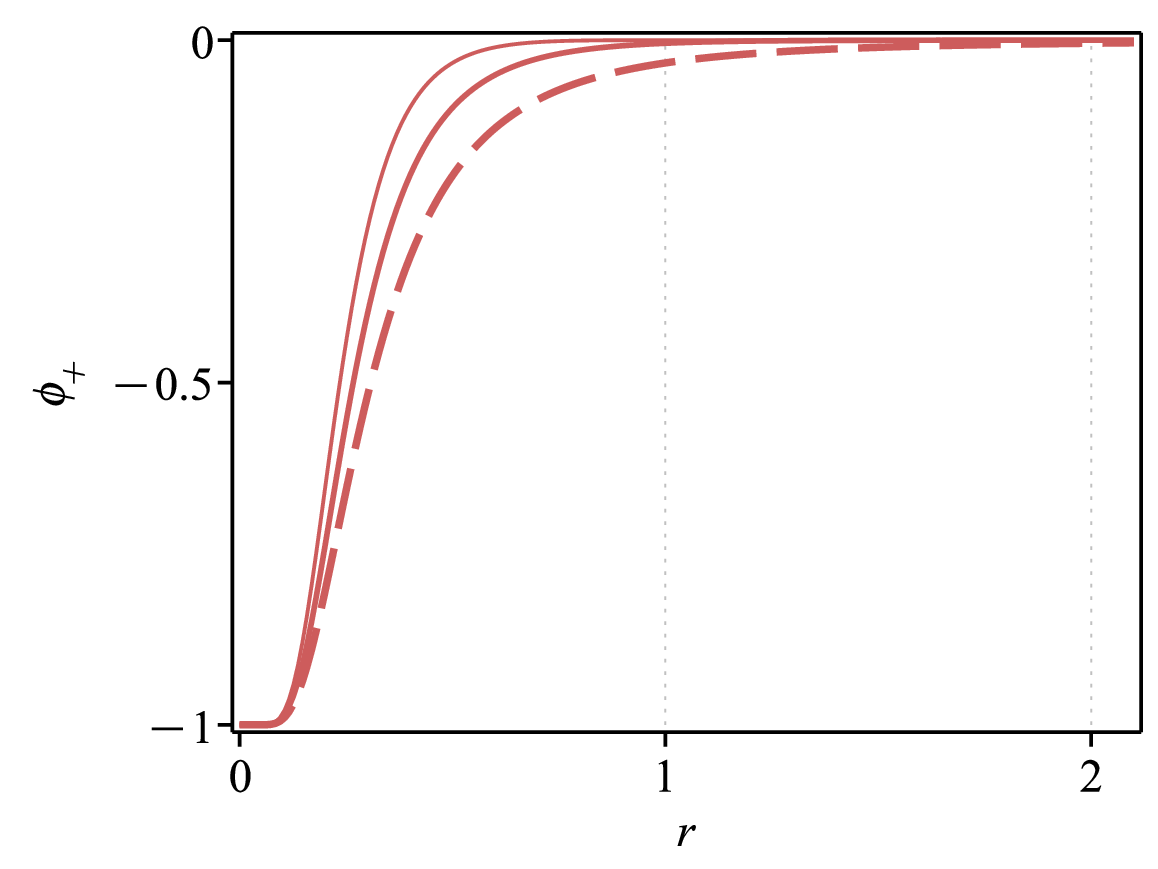}
    \includegraphics[width=4.2cm]{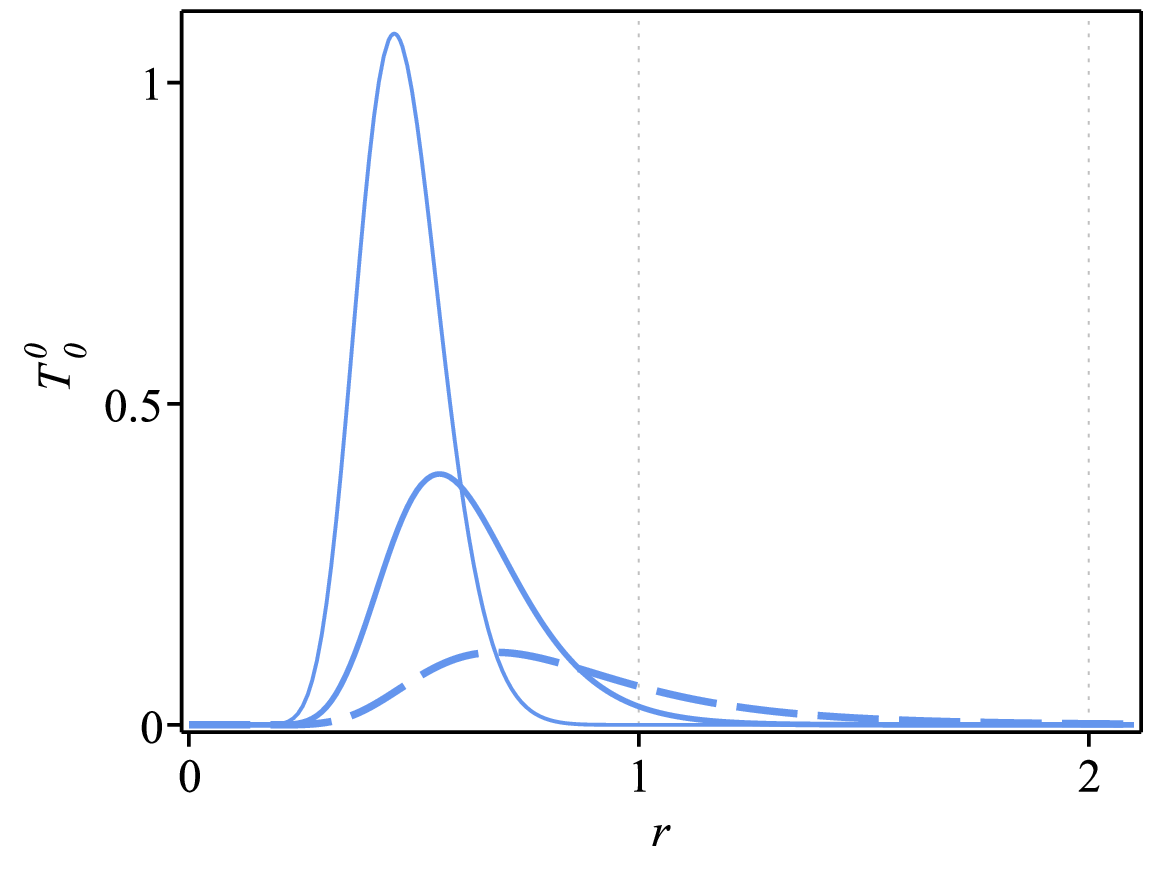}
    \includegraphics[width=4.2cm]{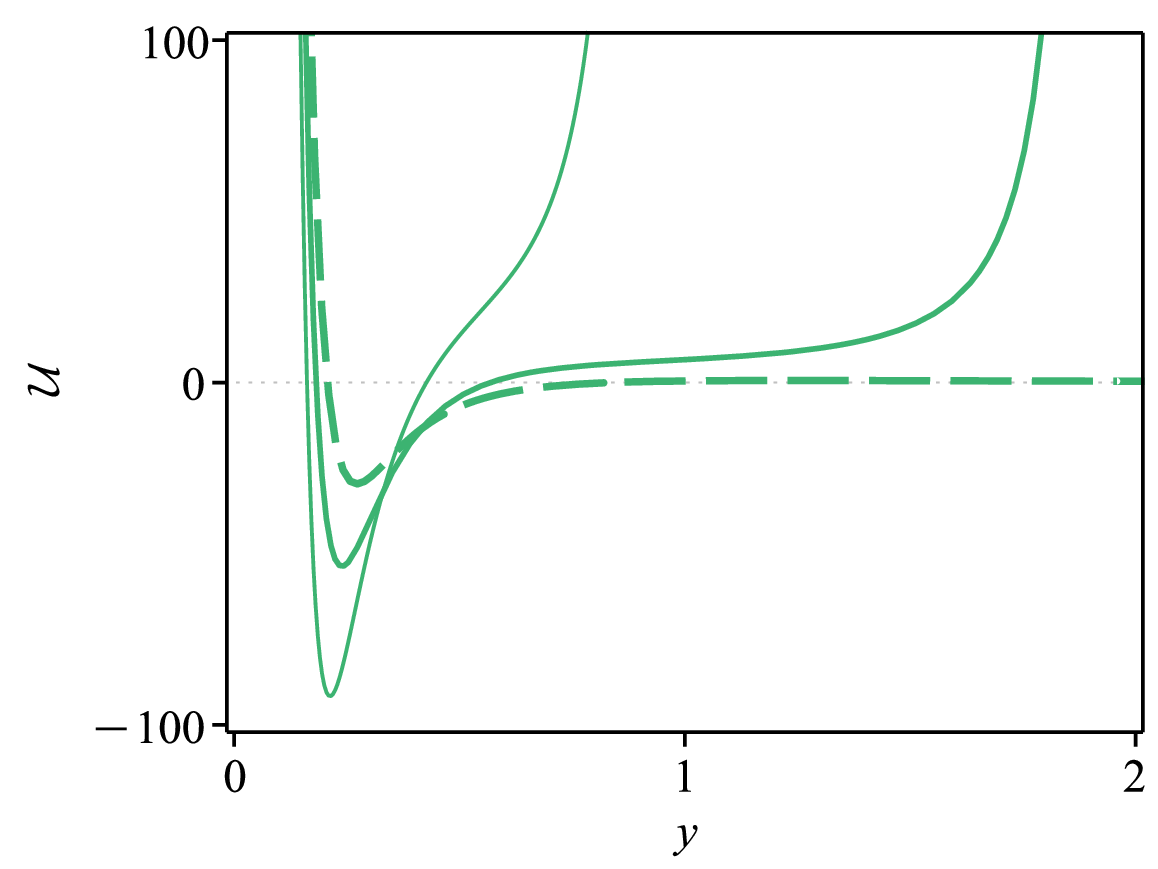}
    \caption{The charge density $\varrho(r)$ in Eq.~\eqref{rhoexp} (top left, orange) for $e=1$, $\alpha=0.01, 0.05, 0.1$ with the thickness of the lines increasing with $\alpha$. We also display the solution $\phi_+(r)$ in Eq.~\eqref{solexp} (top right, red), energy density $\tensor{T}{^0_0}(r)$ in Eq.~\eqref{t00exp} (bottom left, blue) and Schr\"odinger-like potential $\Uc(y)$ in Eq.~\eqref{Uyexp} (bottom right, green), for $p=3$, $\alpha=0.01$, $r_c=1,2$ and $r_c\to\infty$, with the thickness of the lines now increasing with $r_c$. The dashed lines represent the limit $r_c\to\infty$ in each quantity and the vertical dotted lines in gray stand for the value $r=r_c$.}
    \label{fig2}
\end{figure}

Let us now investigate the exponential metric in isotropic coordinates \cite{expmetric1,expmetric2,expmetric3}, with line element
\be\label{ds2exp}
ds^2 = \textrm{e}^{\alpha/r} dt^2-\textrm{e}^{-\alpha/r}\left(dr^2+r^2d\theta^2 + r^2\sin^2\theta d\varphi^2\right),
\ee
where $\alpha$ is a real parameter. In this case, we have $\sqrt{|g|}=r^2\textrm{e}^{-\alpha/r}\sin\theta$ and $\sqrt{\tau}=r^2\textrm{e}^{-\alpha/r}$. To use the first-order framework, we must obey the conditions in Eq.~\eqref{condbps}, which implies that $\Qc(r) = e\,\textrm{e}^{-\alpha/(2r)}$. Since our procedure requires that the above expression vanishes at the origin, we impose that $\alpha$ be strictly positive.
The charge density in Eq.~\eqref{densQbps} can be written as
\be\label{rhoexp}
\varrho(r) = \frac{\alpha e}{2r^4}\,\textrm{e}^{\alpha/(2r)},
\ee
which is positive in all the space. We display it the top-left panel of Fig.~\ref{fig2}. By integrating the above equation, we get the finite charge $Q = 4\pi e$, matching with the value obtained from Eq.~\eqref{Qbps}, being independent on $\alpha$.

To obtain the scalar field profile, we take advantage of the change of variables \eqref{dx}, which leads us to $x = 1/r_c -1/r$. The parameter $r_c$ separates two regions in the mapping. The region $r\in(r_c,\infty)$ corresponds to $x\in(0,1/r_c)$, which is a compact space, so it is not compatible with the solutions \eqref{solp}. The other region is described by the interval $r\in(0,r_c)$, which is mapped into $x\in(-\infty, 0)$. Since $x$ is a semi-compact space, we can use the solution in Eq.~\eqref{solp1}, which becomes
\be\label{solexp}
\phi_\pm(r) = \begin{cases}
    \pm\tanh^p\left(\cfrac{r-r_c}{p\,r\,r_c}\right), & r\leq r_c\\
    0, &r>r_c,
\end{cases}
\ee
where $r_c$ denotes the compactification radius and the upper/lower sign represents the increasing/decreasing solution. Notice that the compact support is lost for $r_c\to\infty$, for which the solution becomes extended, as usually occurs in kinks. In the top-right panel of Fig.~\ref{fig2} one can see the behavior of $\phi_+(r)$ for some values of the parameters. The energy density in Eq.~\eqref{t00bps} is also compact; it reads
\be\label{t00exp}
\tensor{T}{^0_0}(r) = \frac{1}{r^4}\,\textrm{e}^{\alpha/r}\tanh^{2(p-1)}\left(\frac{r-r_c}{p\,r\,r_c}\right)\sech^4\left(\frac{r-r_c}{p\,r\,r_c}\right),
\ee
for $r\leq r_c$ and $\tensor{T}{^0_0}(r)=0$ otherwise. By integrating it, the energy is $\Ec=2p\Omega(d)/(4p^2-1)$, as expected from the definition of $\Ec_B$ above Eq.~\eqref{fo}. The behavior of the energy density at the origin depends on $\alpha$: it is null for $\alpha<4$ and infinite otherwise. We display this energy density in the bottom-left panel of Fig.~\ref{fig2}.

Let us now focus on the stability of the solution \eqref{solexp} around small fluctuations. The functions \eqref{sigmasl} and \eqref{ssl} which describes the Sturm-Liouville operator \eqref{stab} are
\be
\sigma(r) = r^2\textrm{e}^{-2\alpha/r}\quad\text{and}\quad
s(r)= \textrm{e}^{\alpha/r},
\ee
from which we see that $\sigma(r)$ vanishes at the origin and diverges asymptotically, and $s(0)\to\infty$ and $s(\infty)\to1$. The aforementioned operator also depends on the stability potential in Eq.~\eqref{usl}, which can be written in the form
\begin{equation}
\begin{aligned}
U(r) &= \frac{1}{r^4}\textrm{e}^{2\alpha/r}\left(\left(1+\frac1p\right)\left(1+\frac2p\right)\tanh\left(\frac{r-r_c}{p\,r\,r_c}\right)^2 \right.\\
&\left.+\left(1-\frac1p\right)\left(1-\frac2p\right)\tanh\left(\frac{r-r_c}{p\,r\,r_c}\right)^{-2} -2\right),
\end{aligned}  
\end{equation}
for $r\leq r_c$, and $U(r)\to\infty$ for $r>r_c$, if $r_c$ is finite. The case in which $r_c\to\infty$ obeys to $U(\infty)\to0$. The eigenvalue equation that describes the linear stability admits the zero mode in Eq.~\eqref{zeromode}, which leads us to
\be\label{zeromodeexp}
\eta_0(r) = \Nc\tanh^{p-1}\left(\frac{r-r_c}{p\,r\,r_c}\right)\sech^2\left(\frac{r-r_c}{p\,r\,r_c}\right)
\ee
for $r\leq r_c$, and $\eta_0(r)=0$ for $r>r_c$. This zero mode can be normalized, obeying Eq.~\eqref{condnorm}. Unfortunately, we were not able to find a general expression for $\Nc$, so it must be calculated for each set of values of the parameters. It is worth remarking that, as a consequence of our first-order framework, the adjoint operators \eqref{selfadjoint} that factorize the Sturm-Liouville operator are given by
\bes
\bal
S&= -\textrm{e}^{\alpha/r}\Bigg(\frac{d}{dr} +\frac{1}{r^2}\bigg(\left(\frac1p +1\right)\tanh\left(\frac{r-r_c}{p\,r\,r_c}\right)\nn
    &+\left(\frac1p-1\right)\tanh^{-1}\left(\frac{r-r_c}{p\,r\,r_c}\right)\bigg)\Bigg)\\
S^\dagger &= \textrm{e}^{\alpha/r}\Bigg(\frac{d}{dr} -\frac{1}{r^2}\bigg(\left(\frac1p +1\right)\tanh\left(\frac{r-r_c}{p\,r\,r_c}\right)\nn
    &+\left(\frac1p-1\right)\tanh^{-1}\left(\frac{r-r_c}{p\,r\,r_c}\right)\bigg) +\frac{\alpha+2r}{r^2}\Bigg)
\eal
\ees
for $r\leq r_c$. These operators are not well defined for $r>r_c$ as the stability potential is infinite in this interval.

As we have commented in the general study of the stability, one can perform the change of variables \eqref{change} to transform the Sturm-Liouville equation into a Schr\"odinger-like one. For the metric under investigation, we get
\be\label{yrexp}
y = r\textrm{e}^{-\alpha/r} +\alpha\text{Ei}\left(-\frac{\alpha}{r}\right),
\ee
where Ei$(z)$ is the exponential integral function with argument $z$. The above equation shows that its asymptotic behavior is $y\approx r$. However, it is not possible to write $r$ in terms of $y$ analytically, so numerical procedures are required if one wishes to use the Schr\"odinger-like equation. In particular, the Schr\"odinger-like potential in Eq.~\eqref{hu} is given by
\be\label{Uyexp}
\Uc(y) = U(r(y)) -\frac{\alpha\textrm{e}^{-\alpha/(2r(y))}}{4r(y)^4}(\alpha+4r(y))
\ee
for $y\leq y_c\equiv r_c\textrm{e}^{-\alpha/r_c} +\alpha\text{Ei}(-\alpha/r_c)$, and $\Uc(y)\to\infty$ for $y>y_c$. We have used $r(y)$ to denote the inverse of Eq.~\eqref{yrexp}. The above expression leads us to conclude that the above potential is an infinite well for finite $y_c$. This behavior, however, is changed in the limit $y_c\to\infty$, for which we get $\Uc(y)\approx (p-1)(p-2)y^{-2}$, so the potential vanishes for large $y$. In the bottom-right panel of Fig.~\ref{fig2} we display the above Schr\"odinger-like potential. Even though numerical methods are required, we can see that the zero mode \eqref{zeromodeexp} does not engender nodes; this ensures that negative eigenvalue is absent and the model is linearly stable.

In Ref. \cite{expmetric3}, the authors have found interesting connection between the exponential metric considered above with a traversable wormhole, so the above study may motivate other investigations, adding dynamics to the geometric degrees of freedom, searching for the possibility to find a more general framework engendering first-order differential equations.  

\subsection{Hyperscaling violating geometries}

\begin{figure}[t!]
    \centering
    \includegraphics[width=6cm]{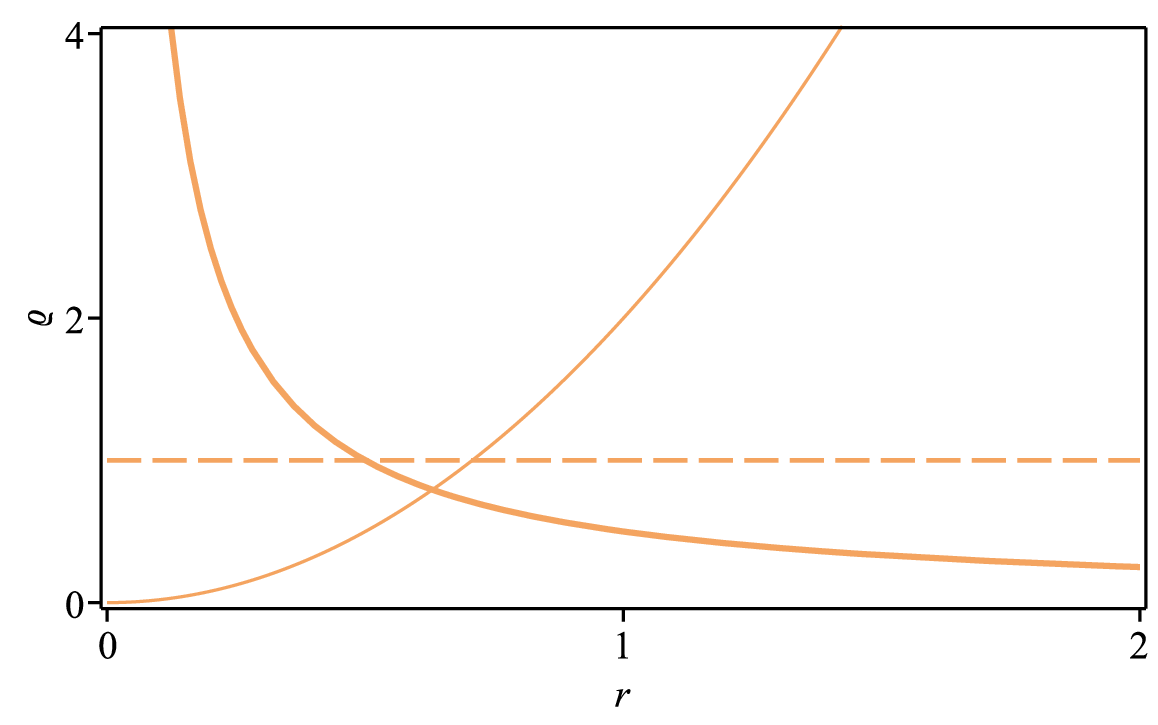}
    \caption{The charge density $\varrho(r)$ in Eq.~\eqref{rhoexplifs} for $e=1$, $\theta_c=0$ and $z=-2,-1,-1/2$. The thickness of the lines increases with $z$. The dashed line represents $z=-1$, which is the case where the charge density $\varrho(r)$ is constant.}
    \label{fig3}
\end{figure}

We consider the metric associated to hyperscaling violating geometries \cite{app1} in $(d,1)$ spacetime dimensions, with $\omega_{ij}=\delta_{ij}$ and $\theta^i=x^i$, defined by the line element
\be
ds^2 = r^{2z-\theta_c}dt^2 -r^{-2-\theta_c}dr^2 -r^{2-\theta_c}dx^kdx^k.
\ee
The above expression leads to $\sqrt{|g|}=\sqrt{\tau}=r^{d+z-2-(\theta_c/2)(d+1)}$ and $k=1,2,...,d-1$, in which $z$ and $\theta_c$ represents the dynamical and hyperscaling violating exponents, respectively. For the value $\theta_c=0$, the Lifshitz spacetime \cite{lif} is recovered. To work with the first-order framework, we consider the condition in the left equation of \eqref{condbps}, which requires $\Qc(r) = er^{\theta_c/2-z}$. To avoid divergences in this function at the origin, we impose $\Qc(0)=0$, which is satisfied by $\theta_c>2z$. In this case, the charge density \eqref{densQbps} associated to the current geometric background reads
\be\label{rhoexplifs}
\varrho(r) = \frac{e}{2}(\theta_c-2z)r^{(\theta_c/2)(d+2)-d-2z+1}.
\ee
It is regular, without divergence at the origin, if $\theta_c\geq 2(d+2z-1)/(d+2)$. It is always non negative and leads to infinite charge, as expected from Eq.~\eqref{Qbps}. In Fig.~\ref{fig3}, we display the behavior of the above charge density for some values of the parameters, showing that it may be null, finite or divergent both at the origin or asymptotically.

\begin{figure}[t!]
    \centering
     \includegraphics[width=4.2cm]{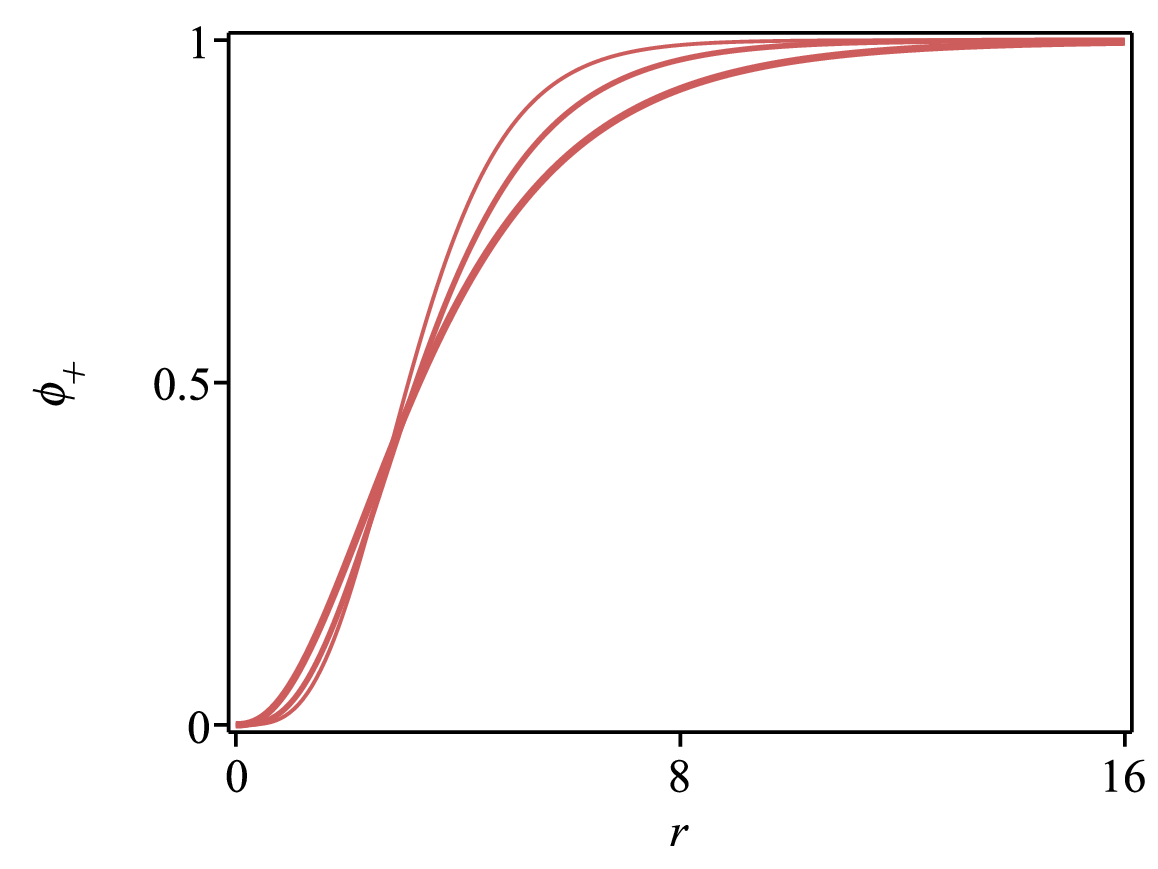}
    \includegraphics[width=4.2cm]{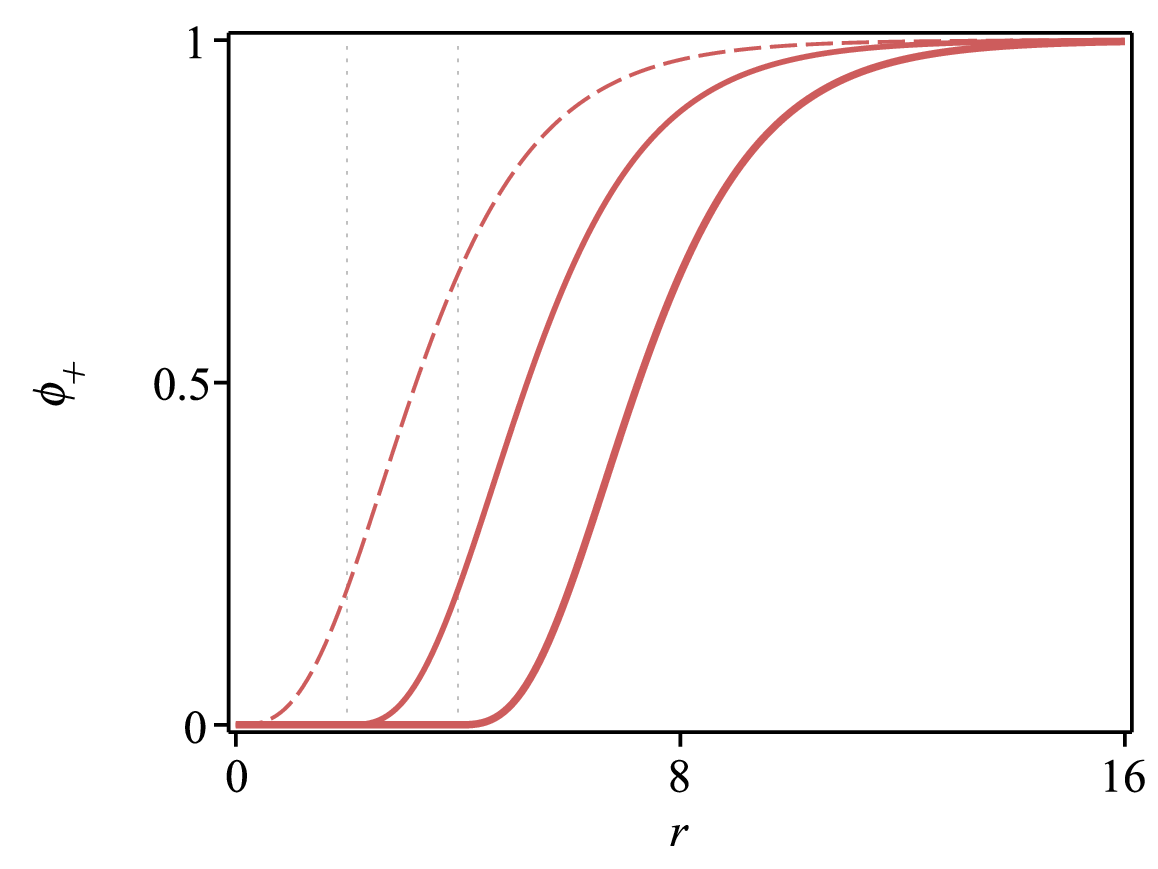}
    \includegraphics[width=4.2cm]{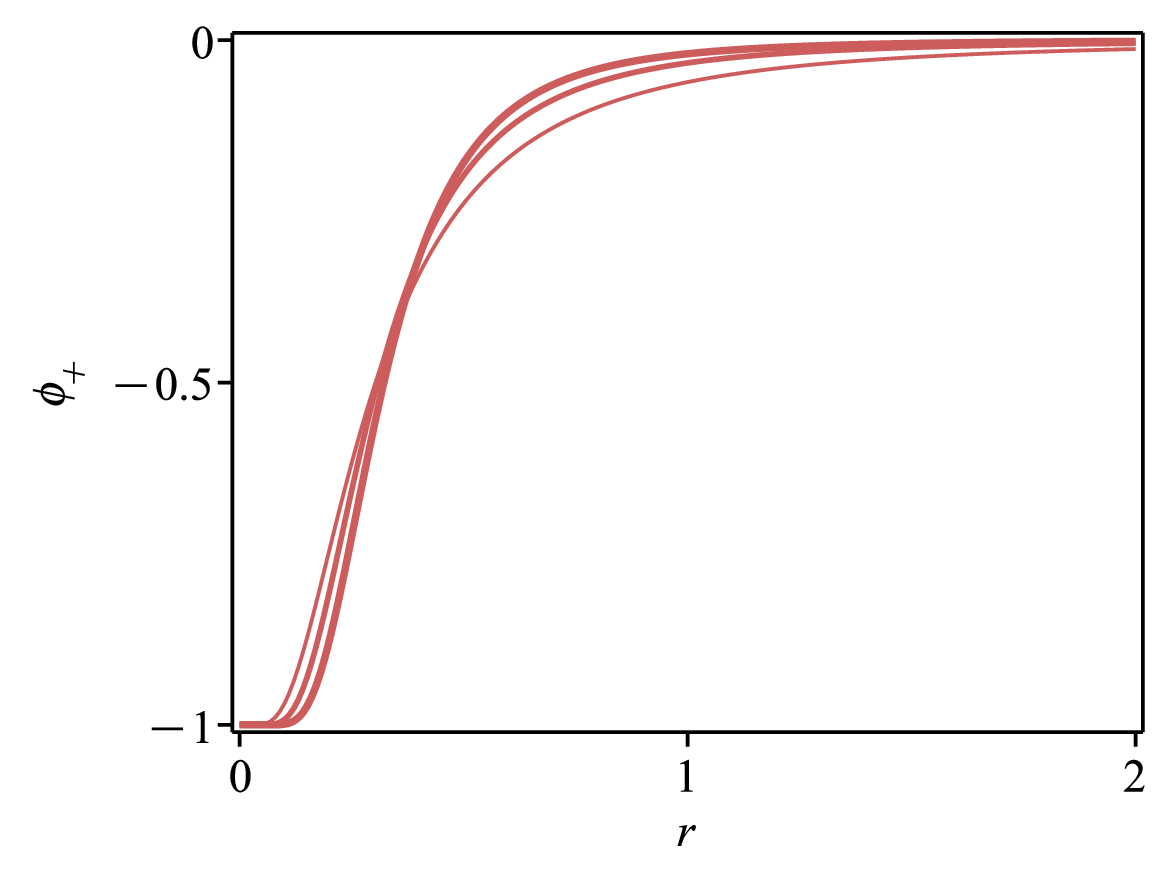}
    \includegraphics[width=4.2cm]{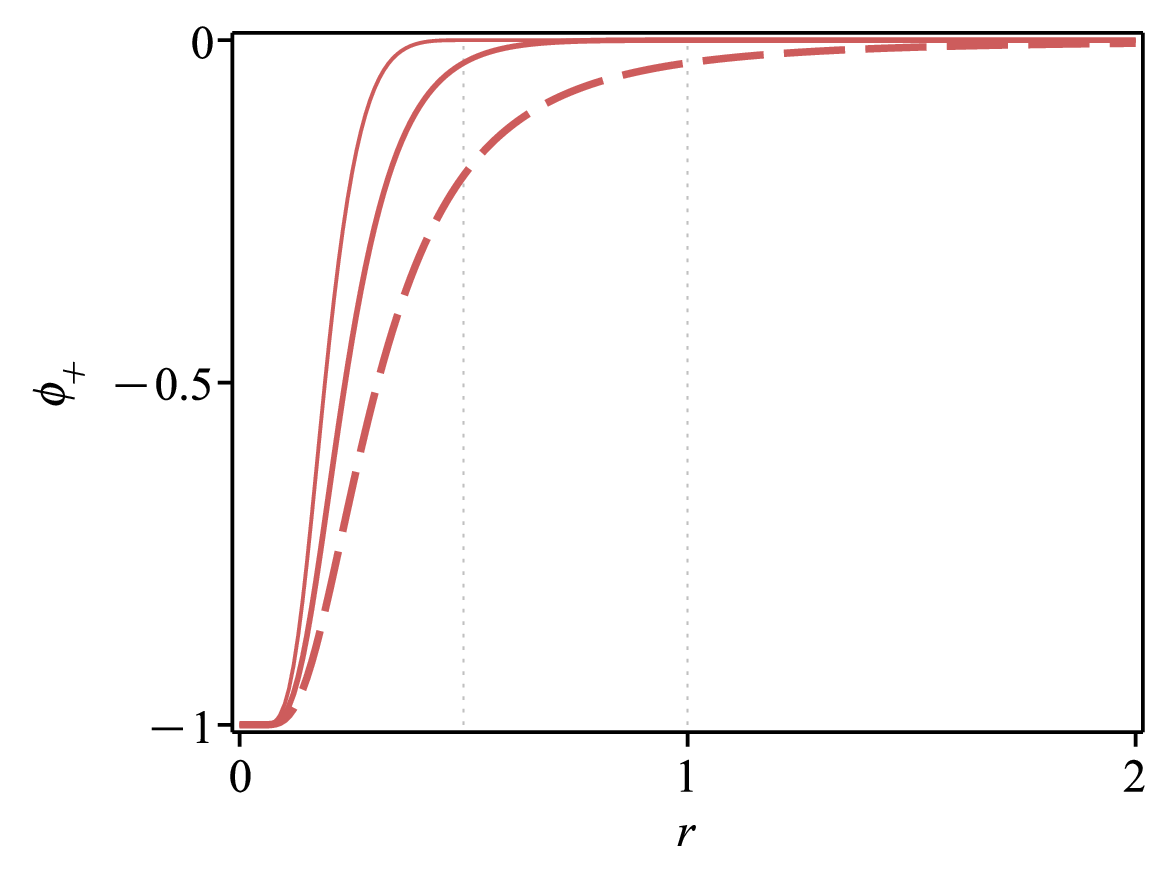}
    \caption{The solution $\phi_+(r)$ in Eq.~\eqref{solhypern}. In the top (bottom) panels, we depict the case $n>0$ ($n<0$), for $p=3$ and $\theta_c=0$. In the top-left panel, we use $r_c=0$ and $z=-3.2,-3$ and $-2.8$. In the top-right panel, we take $z=-3$ and $r_c=0,2$ and $4$. In the bottom-left panel, we consider $r_c\to\infty$ and $z=-1.2,-1$ and $-0.8$. In the bottom-right panel, we have $z=-1$ and $r_c=1/2,1$ and $r_c\to\infty$. The dashed lines represent the case $r_c=0$ in the top-right panel and $r_c\to\infty$ in the bottom-right panel. The dotted vertical lines represent the point $r=r_c$ in the solutions. The thickness of the lines increases with $z$ in the left panels and with $r_c$ in the right ones.}
    \label{fig4}
\end{figure}

Since we are considering the model described by \eqref{wpmodel}, the solution is given by Eq.~\eqref{solp}, with argument $x$ related to $r$ via Eq.~\eqref{dx}, which leads us to $x = \big(r^n-r_c^n\big)/n$, where $n = \theta_c(d-1)/2-d-z+1$ and $r_c$ is an integration constant which determines the point in which $x=0$. Notice that the case $n=0$ is special; we shall investigate it later. For non-null $n$, one can use the solutions \eqref{solp} to show that
\be\label{solhypern}
\phi_\pm(r) \!=\! \begin{cases}\pm\tanh^p\left(\frac{1}{np}\big(r^n-r_c^n\big)\right),\!\! & r\,\text{sgn}(n)\geq r_c\,\text{sgn}(n)\\
0,\!\! & r\,\text{sgn}(n)< r_c\,\text{sgn}(n),
\end{cases}
\ee
where ${\rm sgn}(x)$ denotes the signal function. The upper/lower sign represents the increasing/decreasing solution. The solution $\phi_+(r)$ is depicted in Fig.~\ref{fig4} for positive and negative values of $n$ and some values of the parameters.

\begin{figure}[t!]
    \centering
    \includegraphics[width=4.2cm]{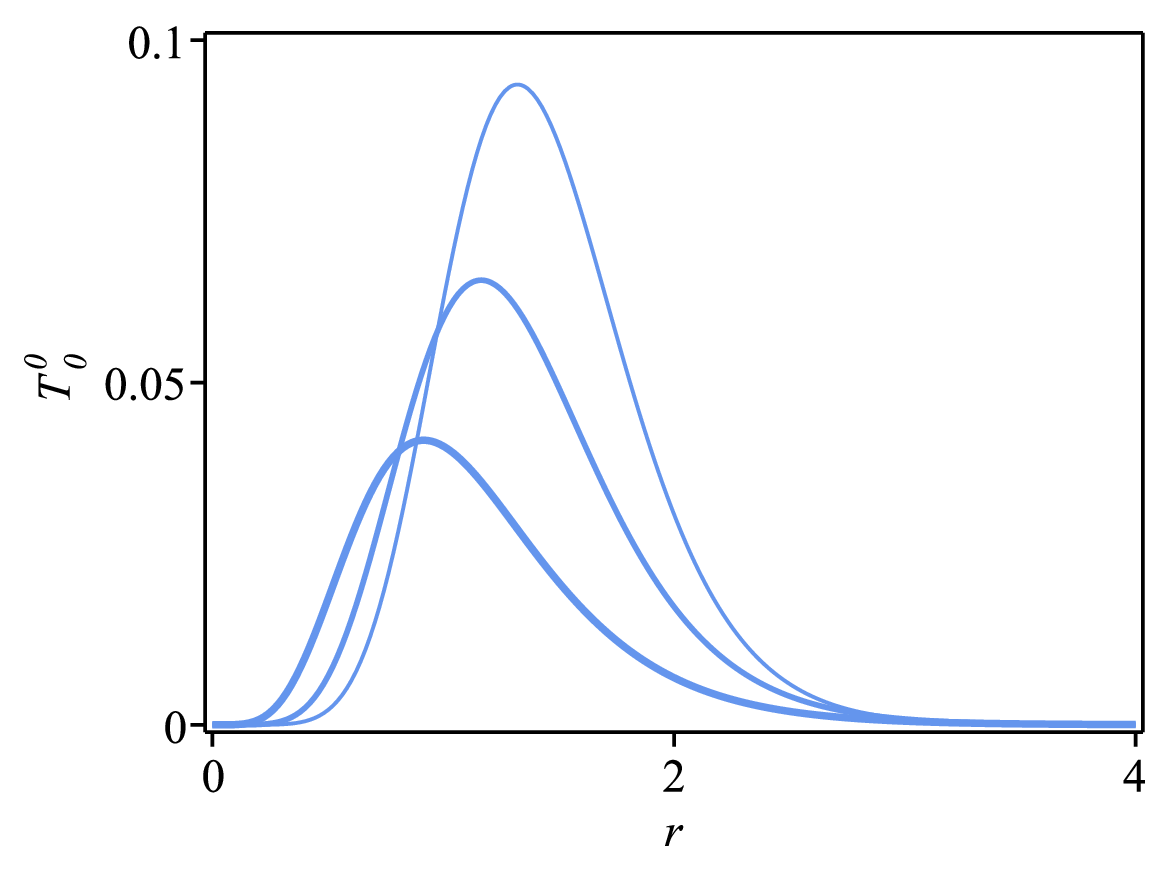}
    \includegraphics[width=4.2cm]{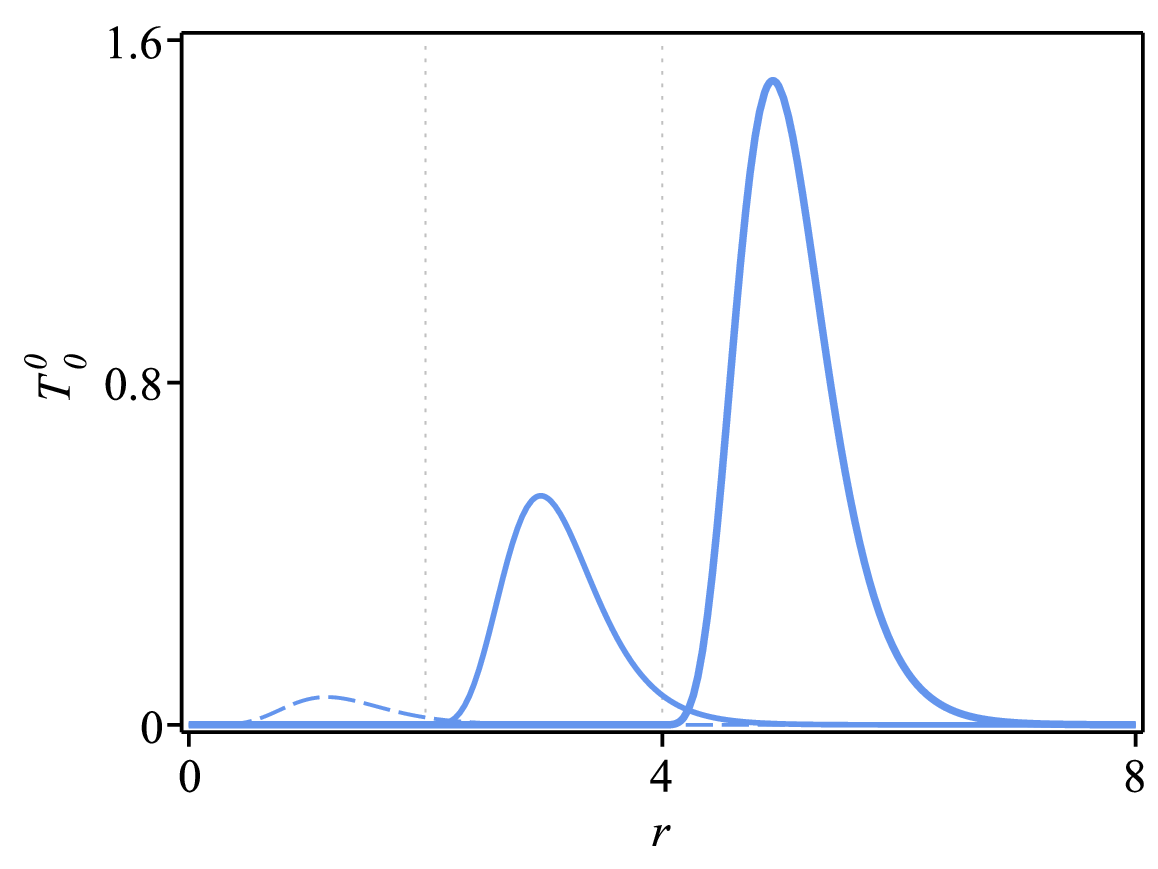}
    \includegraphics[width=4.2cm]{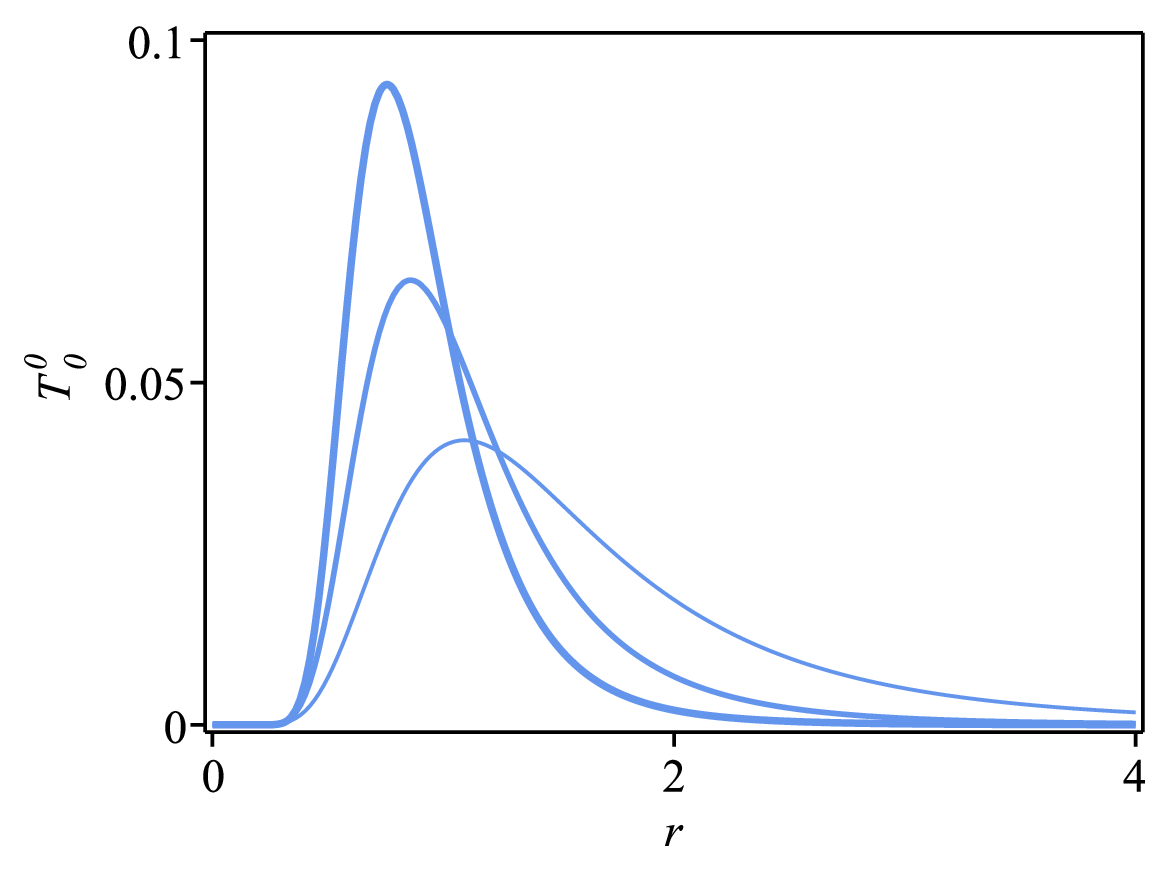}
    \includegraphics[width=4.2cm]{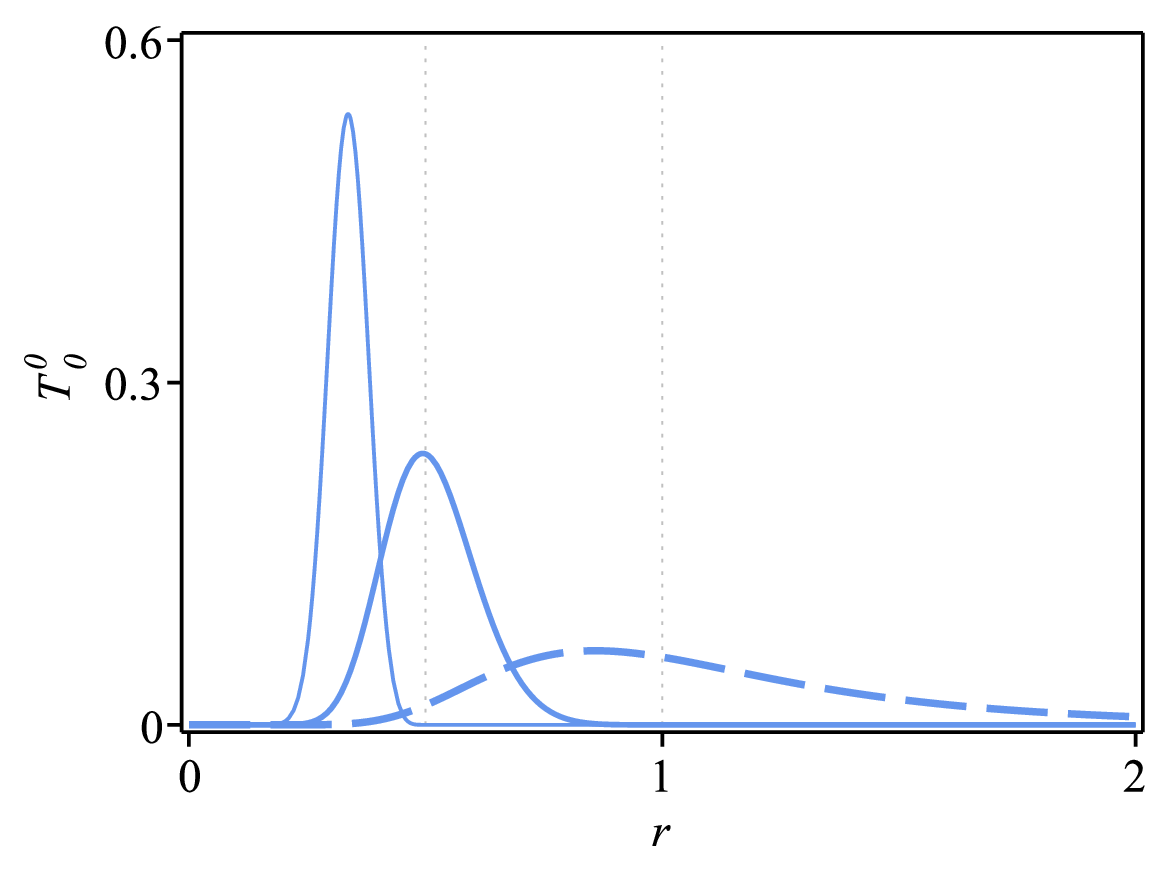}
    \caption{The energy density $\tensor{T}{^0_0}(r)$ in Eq.~\eqref{t00hypn}. In the top (bottom) panels, we depict the case $n>0$ ($n<0$), for $p=3$ and $\theta_c=0$. In the top-left panel, we use $r_c=0$ and $z=-3.2,-3$ and $-2.8$. In the top-right panel, we take $z=-3$ and $r_c=0,2$ and $4$. In the bottom-left panel, we consider $r_c\to\infty$ and $z=-1.2,-1$ and $-0.8$. In the bottom-right panel, we have $z=-1$ and $r_c=1/2,1$ and $r_c\to\infty$. The dashed lines represents the case $r_c=0$ in the top-right panel and $r_c\to\infty$ in the bottom-right panel. The dotted vertical lines represent the point $r=r_c$. The thickness of the lines increases with $z$ in the left panels and with $r_c$ in the right ones.}
    \label{fig5}
\end{figure}

For $n>0$, the solution $\phi_+$ ($\phi_-)$ is uniform and null in the interval $r<r_c$, and connects $\phi=0$ to $\phi=1$ ($\phi=-1$), in $r\geq r_c$. The case $n<0$ is interesting, as it supports compact solutions for finite $r_c$. In this situation, $\phi_+$ connects $\phi=-1$ to $\phi=0$ and $\phi_-$ goes from $\phi=1$ to $\phi=0$. The limit $r_c\to\infty$ decompactifies the solution, leading to $\phi_\pm= \pm\tanh^p\left(r^n/(np)\right)$. The energy density \eqref{t00bps} associated to the above solutions becomes
\be\label{t00hypn}
\begin{aligned}
\tensor{T}{^0_0}(r) &= r^{2n+\theta_c}\tanh\left(\frac{1}{np}\big(r^n-r_c^n\big)\right)^{2p-2}\\
    &\times\sech\left(\frac{1}{np}\big(r^n-r_c^n\big)\right)^4
\end{aligned}
\ee
for the regions in which the solution \eqref{solhypern} is not uniform, and $\tensor{T}{^0_0}(r)=0$ otherwise. For positive $n$, as we approach $r=r_c$ from the right, one can show that the above expression behaves as $\tensor{T}{^0_0}(r)\propto r^{2n+\theta_c}(r^n-r^n_c)^{2(p-1)}$. On one hand, this shows that $\tensor{T}{^0_0}(r)$ is always null at $r=r_c$ for $r_c>0$. On the other hand, for $r_c=0$, the behavior can be null, finite or divergent at $r=0$, depending on the sign of $2pn+\theta_c$. For negative $n$, the behavior near the origin has the form $\tensor{T}{^0_0}(r)\propto r^{2n+\theta_c}\,\textrm{e}^{-4r^n}$, so the energy density always vanish at the aforementioned point. Notice that the energy density is compact for negative values of $n$ when $r_c$ is finite, as expected from the solution \eqref{solhypern}. By integrating the above expression, we get the energy $\Ec=2p\Omega(d)/(4p^2-1)$, matching with the value expected from the definition of $\Ec_B$ above Eq.~\eqref{fo}. $\Omega(d)$ represents the Euclidean volume related to the $x^i$-coordinates. The above energy density is plotted in Fig.~\ref{fig5} for some values of the parameters.

\begin{figure}[t!]
    \centering    \includegraphics[width=0.5\linewidth]{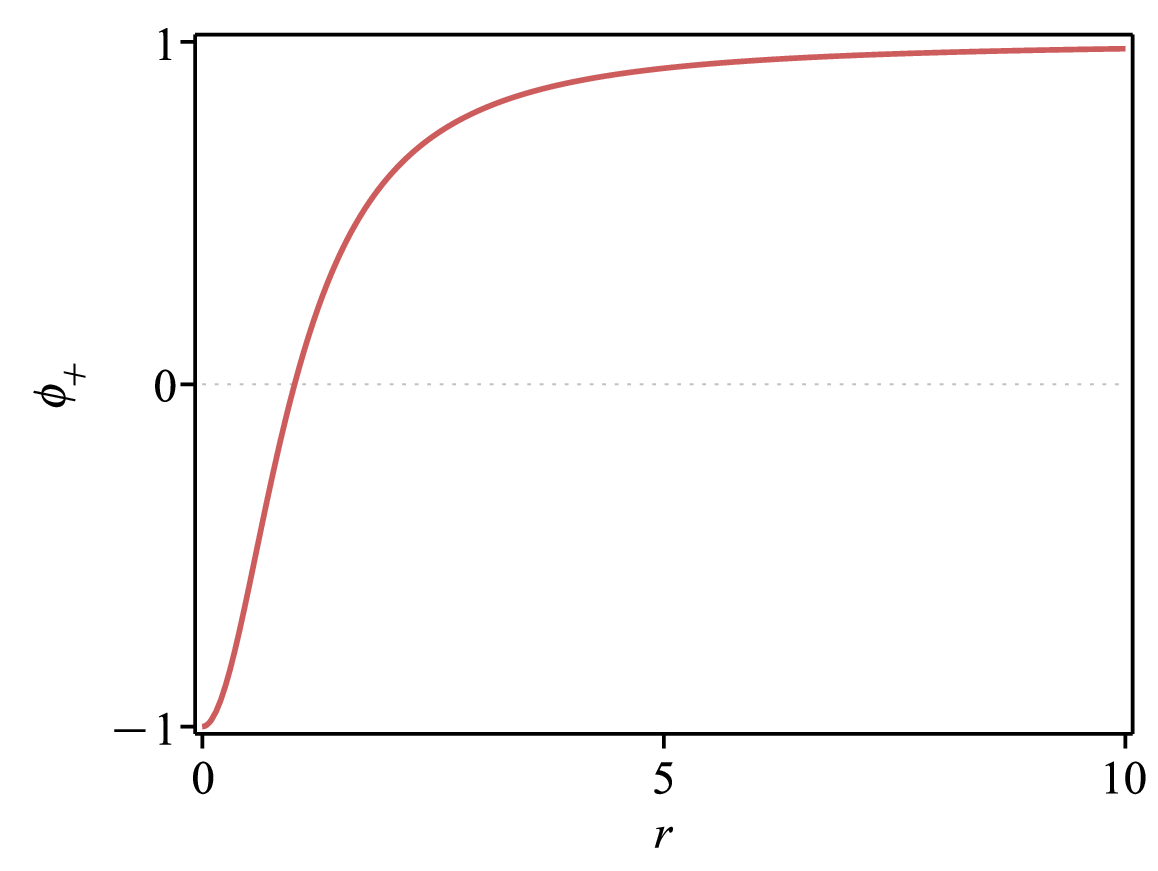}\includegraphics[width=0.5\linewidth]{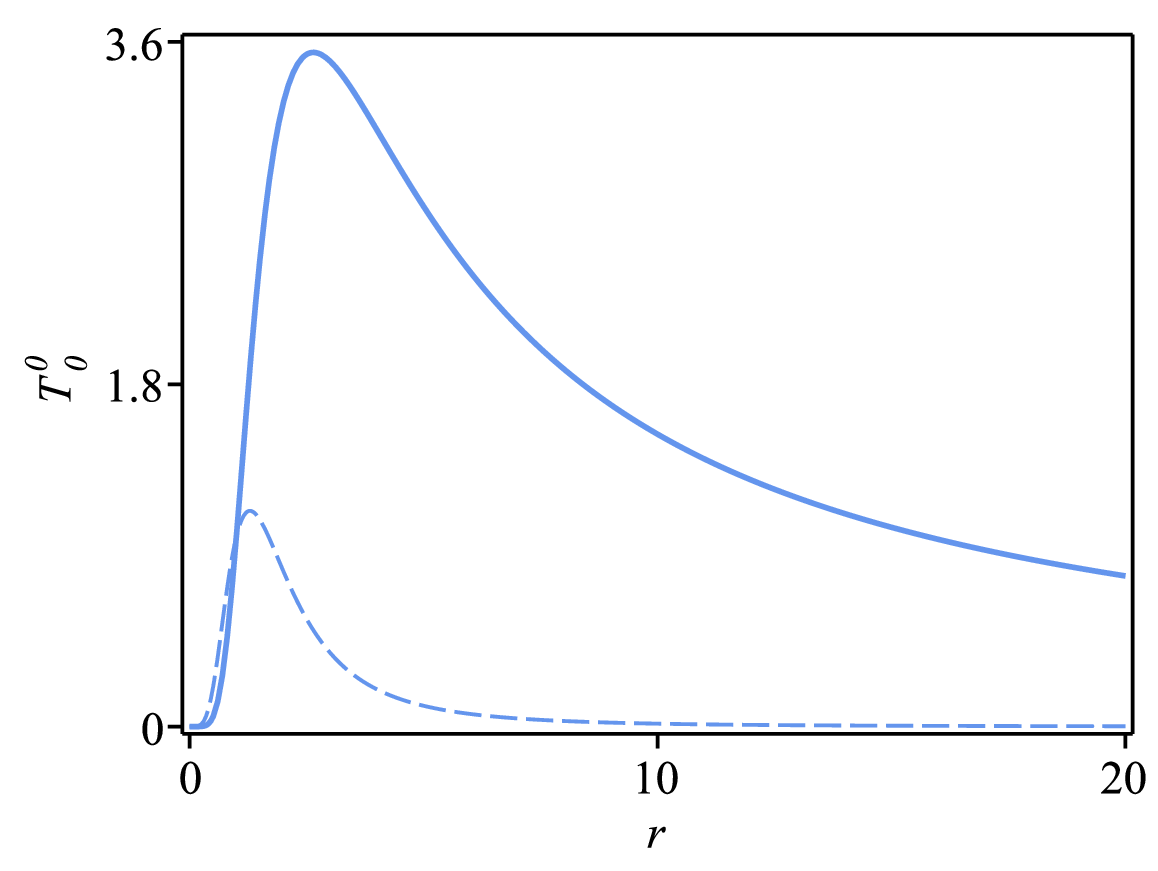}
    \caption{The solution \eqref{solhypn0} (left) and the energy density \eqref{t00hypn0} (right) for $p=1$. In the energy density, the dashed line represents $\theta_c=1$ and $z=-1$, and the solid line stands for $\theta_c=3$ and $z=1$.}
    \label{fig6}
\end{figure}

Let us now deal with the case in which $n=0$, obtained for $\theta_c=2+2z/(d-1)$. The change of variables in Eq.~\eqref{dx} leads us to $x = \ln(r/r_c)$. Since $x$ varies from $-\infty$ to $\infty$, we can use the solution \eqref{solp2kink}, which can now be written in terms of power-law functions, as
\be\label{solhypn0}
\phi_\pm(r) = \pm\left(\frac{r^{2/p}-r_c^{2/p}}{r^{2/p}+r_c^{2/p}}\right)^p.
\ee
This solution connects $\phi=-1$ to $\phi=1$. It is increasing/decreasing  for the upper/lower sign. Interestingly, contrary to the case $n\neq0$, the value $p=1$ is now allowed because the solution connects two divergent points of the permittivity as $r$ goes from $0$ to $\infty$. It is plotted in Fig.~\ref{fig6} for some values of the parameters.
The associated energy density \eqref{t00bps} reads
\be\label{t00hypn0}
\tensor{T}{^0_0}(r) = \frac{16r_c^{4/p}r^{2z/(d-1)+2(2+p)/p}}{\big(r^{2/p}+r_c^{2/p}\big)^4}\left(\frac{r^{2/p}-r_c^{2/p}}{r^{2/p}+r_c^{2/p}}\right)^{2p-2}.
\ee
Near the origin, it behaves as $\tensor{T}{^0_0}(r)\propto r^{2z/(d-1)+2(2+p)/p}$, which may lead to null, finite or divergent behavior at this point. We can see the energy density in Fig.~\ref{fig6}. The integral of the above expression leads to the energy $\Ec=4p\Omega(d)/(4p^2-1)$, as expected from the first-order framework.

We now turn our attention to the stability. The Sturm-Liouville operator in Eq.~\eqref{stab} that governs the eigenvalue equation is now described by the functions \eqref{sigmasl} and \eqref{ssl},
\be
\sigma(r) = r^{-n-2z-1}\quad\text{and}\quad
s(r) = r^{1+z}.
\ee
These expressions are valid for any value of $n$, as they only depend on the factors of the metric. However, we must be careful to calculate the stability potential \eqref{usl} due to the form of the solutions. In the case $n\neq0$, it is given by
\begin{equation}
\begin{aligned}
U(r) &= r^{2(n+z)}\left(\left(1+\frac1p\right)\left(1+\frac2p\right)\tanh^2\left(\cfrac{r^n-r_c^n}{np}\right) \right.\\
&\left.+\left(1-\frac1p\right)\left(1-\frac2p\right)\tanh^{-2}\left(\cfrac{r^n-r_c^n}{np}\right) -2\right),
\end{aligned} 
\end{equation}
for $r\,\text{sgn}(n)\geq r_c\,\text{sgn}(n)$, and $U(r)\to\infty$ otherwise (except for $r_c=0$ with $n>0$ or $r_c\to\infty$ with $n<0$). The case $r_c=0$ for $n>0$ requires some care, so we expand the above expression near the origin to get $U(r)\propto r^{2z}$, which may give rise to null, finite or divergent behavior, depending on the value of $z$. For $n<0$ with $r_c\to\infty$, the asymptotic behavior is $U(r)\propto r^{4n+2z}$, which may lead to the same three behaviors commented in the case $n>0$, depending on $z$.

In the case $n=0$, Eq.~\eqref{usl} becomes
\be
\begin{aligned}
U(r) &= \frac{4r^{2z}}{p^2\big(r^{4/p}-r_c^{4/p}\big)^2}\Big(r^{8/p}+r_c^{8/p}\\
    &-6p\big(r_c^{2/p}r^{6/p}+r_c^{6/p}r^{2/p}\big)+2(2p^2+3)r_c^{4/p}r^{4/p}\Big).
\end{aligned}
\ee
The zero mode associated to the stability equation can be calculated from Eq.~\eqref{zeromode}. For $n\neq0$, it is
\be\label{eta0lifsn}
\eta_0(r)=\Nc\tanh^{p-1}\left(\cfrac{r^n-r_c^n}{np}\right)\sech^2\left(\cfrac{r^n-r_c^n}{np}\right)
\ee
for $r\,\text{sgn}(n)\geq r_c\,\text{sgn}(n)$, and $\eta_0(r)=0$ otherwise. In the case $n=0$, it has the form
\be\label{eta0lifsn0}
\eta_0(r)=\cfrac{4\Nc r_c^{2/p}r^{2/p}}{\big(r^{2/p}+r_c^{2/p}\big)^2}\left(\cfrac{r^{2/p}-r_c^{2/p}}{r^{2/p}+r_c^{2/p}}\right)^{p-1}.
\ee
In the last two expressions, $\Nc$ is a normalization constant. For $n>0$ with $r_c=0$, we must impose $n\geq(1+2z)/(2p-3)$ to get a normalized zero mode. For $n<0$ with $r_c\to\infty$, the aforementioned condition is changed to $n < 2z/(2p-3)$. There are no additional restrictions on the values of $n$ and $z$ if $r_c$ is a non-null finite value. In the case $n=0$, we must impose $-2/p < z <(4-p)/(2p)$.

Since we are using the first-order framework, the Sturm-Liouville operator \eqref{stab} can be factorized into the product of the adjoint operators \eqref{selfadjoint}. In the case $n\neq0$, they are
\bes
\be
\begin{aligned}
    S &= -r^{z-1}\Bigg(\frac{d}{dr} +r^{n-1}\bigg(\left(\frac1p +1\right)\tanh\left(\frac{r^n-r_c^n}{np}\right) \\
    &+\left(\frac1p-1\right)\tanh^{-1}\left(\frac{r^n-r_c^n}{np}\right)\bigg)\Bigg),
\end{aligned}
\ee
\be
\begin{aligned}
   S^\dagger &= r^{z-1}\Bigg(\frac{d}{dr} -r^{n-1}\bigg(\left(\frac1p +1\right)\tanh\left(\frac{r^n-r_c^n}{np}\right)\\
   &+\left(\frac1p-1\right)\tanh^{-1}\left(\frac{r^n-r_c^n}{np}\right)\bigg) -\frac{n+z}{r}\Bigg),
\end{aligned}
\ee
\ees
for $r\,\text{sgn}(n)\geq r_c\,\text{sgn}(n)$; they are not well defined for $r\,\text{sgn}(n)< r_c\,\text{sgn}(n)$ due to the nature of the potential, which may be infinite in this interval depending on $n,z$ and $r_c$. The case $n=0$ works with
\bes
\bal
S &= -r^{z-1}\left(\frac{d}{dr} +\frac{2\big(r^{4/p}+r_c^{4/p}-2pr_c^{2/p}r^{2/p}\big)}{pr\big(r^{4/p}-r_c^{4/p}\big)}\right),\\
S^\dagger &= r^{z-1}\Bigg(\frac{d}{dr} -\frac{1}{pr\big(r^{4/p}-r_c^{4/p}\big)}\Big((2+pz)r^{4/p}\nn
&+(2-pz)r_c^{4/p}-4pr_c^{2/p}r^{2/p}\Big)\Bigg).
\eal
\ees
The zero mode does not present nodes, ensuring the stability of the solution around small fluctuations. One may also conduct this investigation using a Schr\"odinger-like eigenvalue equation defined by the operator in \eqref{hu}. This can be done with the change of variables in Eq.~\eqref{change}, which leads us to $y = -1/(zr^z)$ for $z\neq0$ and $y=\ln(r)$ and $z=0$. For simplicity, we only explore the case $z\neq0$. The stability potential associated to the Schr\"odinger-like equation is given by
\be\label{Uyn}
\begin{aligned}
\Uc(y) &= (-zy)^{-2(n+z)/z}\left(\left(1+\frac1p\right)\left(1+\frac2p\right)T^2(y) \right.\\
&\left.+\left(1-\frac1p\right)\left(1-\frac2p\right)T^{-2}(y) -2\right) +\frac{n^2-z^2}{4z^2y^2}
\end{aligned}
\ee
in the case with $n\neq0$, where we have used the notation $T(y) = \tanh\left((-zy)^{-n/z}-(-zy_c)^{-n/z}\right)$ and also defined $y_c=-1/(zr_c^z)$. For $n=0$, we have
\be\label{Uyn0}
\begin{aligned}
\Uc(y) &= \frac{4}{p^2z^2y^2\big((-zy)^{-4/(pz)}-(-zy_c)^{4/(pz)}\big)^2}\\
    &\times\Big((-zy)^{-8/(pz)}+(-zy_c)^{-8/(pz)}\\
    &-6p(-zy_c)^{-2/(pz)}(-zy)^{-6/(pz)}\\
    &-6p(-zy_c)^{-6/(pz)}(-zy)^{-2/(pz)}\\
    &+2(2p^2+3)(-zy_c)^{-4/(pz)}(-zy)^{-4/(pz)}\Big) -\frac{1}{4y^2}.
\end{aligned}
\ee
The range of $y$ for $n=0$ depends on $z$; $z>0$ leads to $y\in(-\infty,0]$ and $z<0$ leads to $y\in[0,\infty)$.

\begin{figure}[t!]
    \centering
    \includegraphics[width=4.2cm]{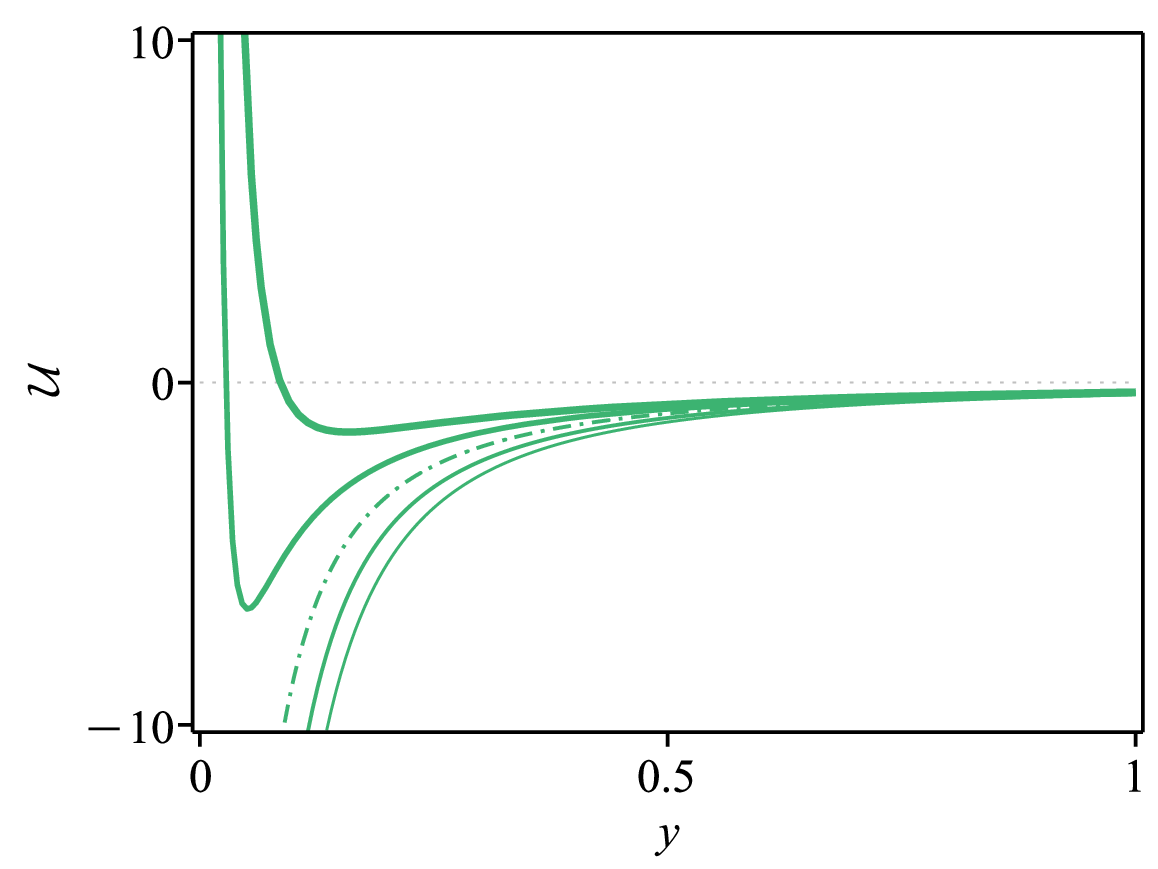}
    \includegraphics[width=4.2cm]{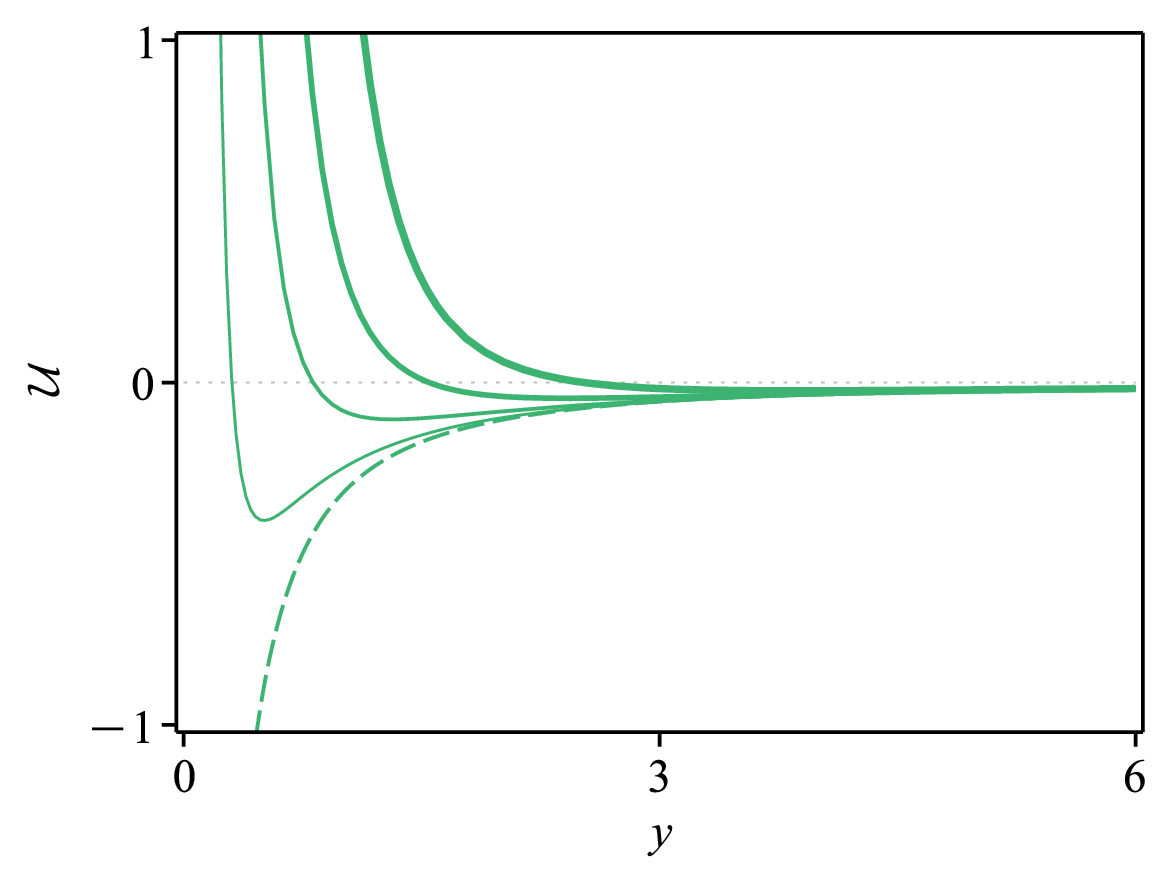}
    \includegraphics[width=4.2cm]{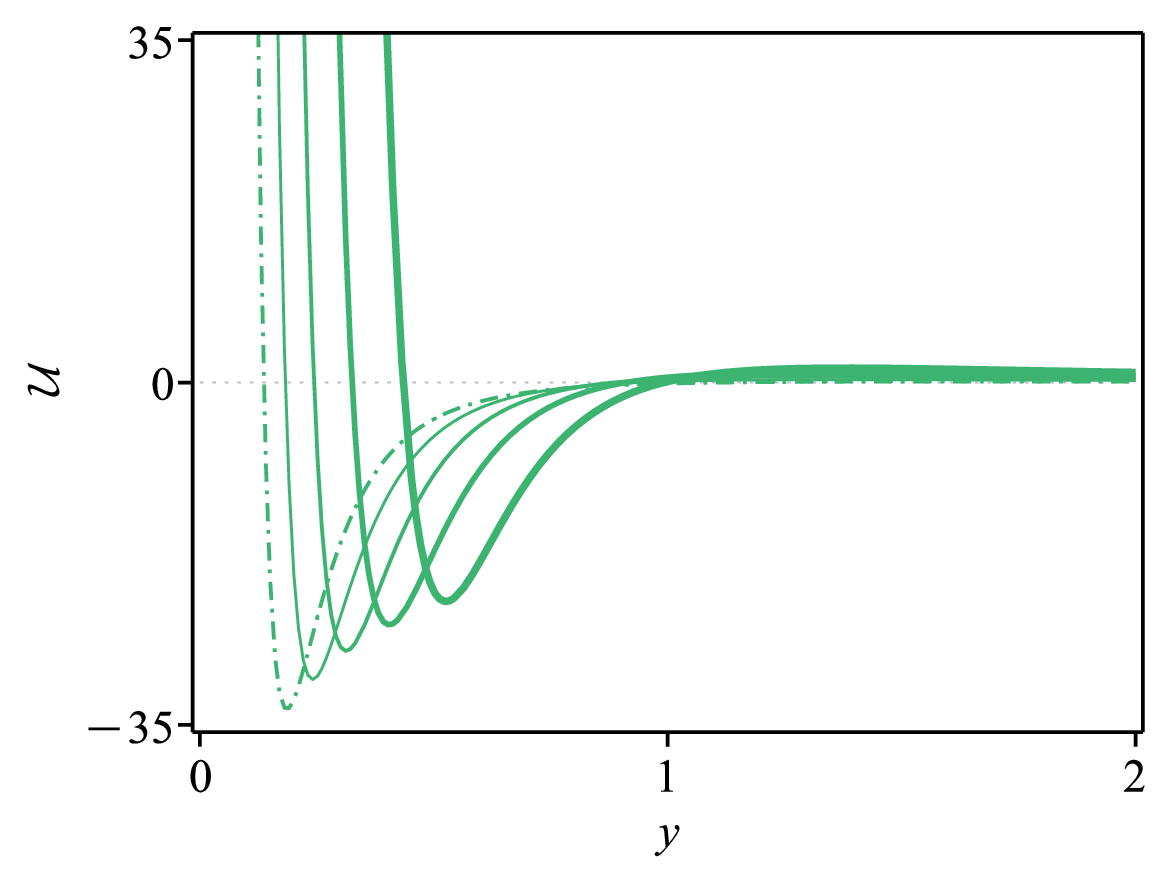}
    \includegraphics[width=4.2cm]{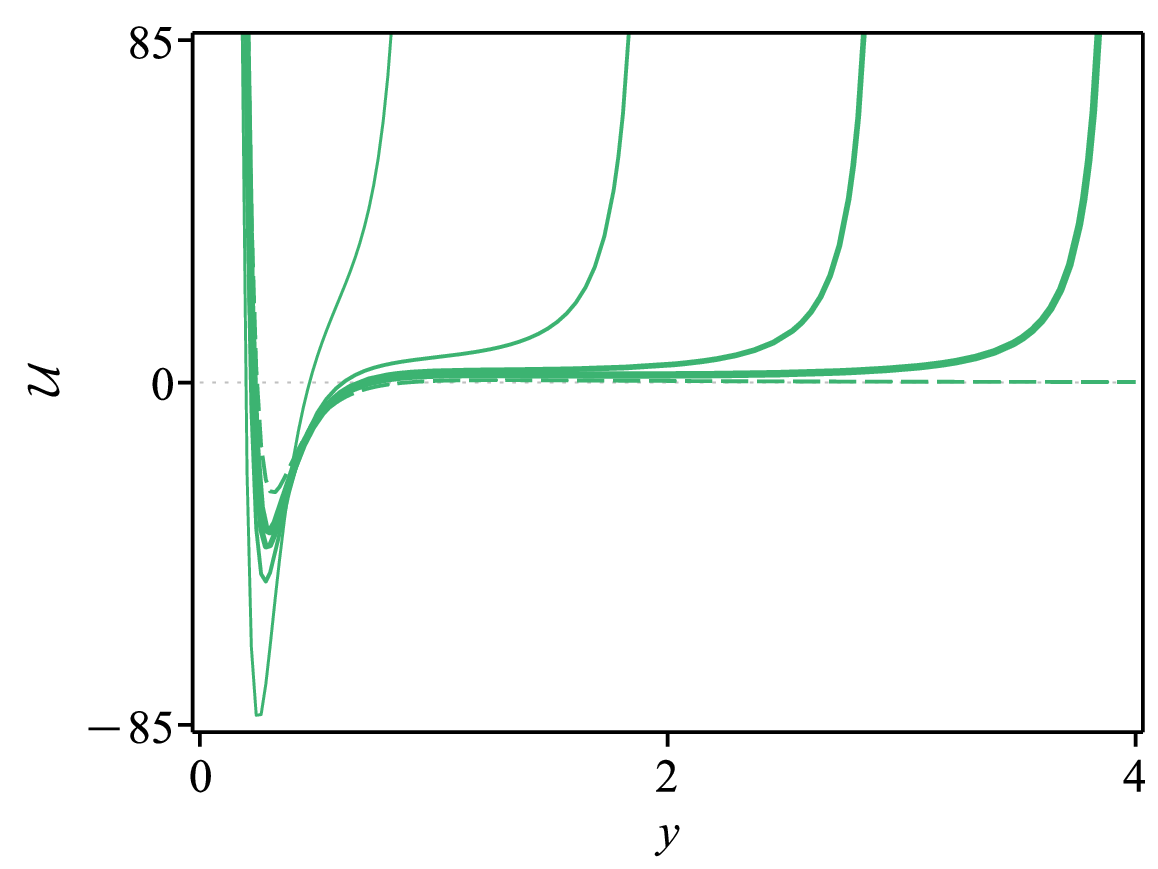}
    \caption{The Schr\"odinger potential $\Uc(y)$ in Eq.~\eqref{Uyn}. In the top (bottom) panels, we depict the case $n>0$ ($n<0$), for $p=3$ and $\theta_c=0$. In the top-left panel, we use $y_c=0$ and $z=-3.2,-3.1,-3,-2.9$ and $-2.8$. In the top-right panel, we take $z=-3$ and $y_c=0,1/192,1/24,9/64$ and $1/3$. In the bottom-left panel, we consider $y_c\to\infty$ and $z=-1.2,-1.1,-1,-0.9$ and $-0.8$. In the bottom-right panel, we have $z=-1$ and $y_c=1,2,3,4$ and $y_c\to\infty$. The dashed lines represent the case $y_c=0$ in the top-right panel and $y_c\to\infty$ in the bottom-right panel. The dash-dotted lines denote the case $z=-3$ in the top-left panel, which delimits the value of $z$ that separates the range supporting negative and positive divergences, and $z=-1.2$ in the bottom-left panel, which delimits the range $z>-1.2$ in which the zero mode can be normalized. The thickness of the lines increases with $z$ in the left panels and with $y_c$ in the right ones.}
    \label{fig7}
\end{figure}

Notice that, for general $n$, $\Uc(y)$ is controlled by $p$, $z$, $d$ and $\theta_c$, as $n = \theta_c(d-1)/2-d-z+1$. This makes the analysis with the general parameters quite intricate. So, to understand the behavior of the Schr\"odinger-like potential, we shall deal first with the situation where $\theta_c=0$, which represents the Lifshitz spacetime, in three spatial dimensions ($d=3)$, with $p=3$. In this case, the condition $\theta_c>2z$ requires that $z$ must be negative; it also obeys $n=-2-z$. For $n>0$ ($n<0$), we must impose $z<-2$ ($-2<z<0$). We display the Schr\"odinger-like potential in Fig.~\ref{fig7} for some values of the parameters. When one considers $n>0$, the analysis at the origin depends on $y_c$, which can only be finite. For $y_c=0$, the potential behaves as $\Uc(y)\approx(2z+3)(z+3)/(z^2y^2)$ near the origin, so it diverges to $-\infty$ ($+\infty$) if $-3< z<-2$ ($z<-3$); the behavior for $z=-3$ is special, with  $\Uc(y)\approx-2/(3y)^{4/3}$, leading to negative divergence at $y=0$. On the other hand, $y_c\neq0$ leads to $\Uc(y)\to+\infty$ for $y\leq y_c$. The asymptotic behavior is $\Uc(y)\approx(z+1)/(z^2y^2)$, which goes to zero as $y$ increases. The case $n<0$ leads to $\Uc(y)=(4/9)(-zy)^{4/z}$ for $y\approx0$, so it presents a positive divergence. For the limit $y_c\to\infty$, the asymptotic behavior is $\Uc(y)\approx(2z+3)(z+3)/(z^2y^2)$, which tends to vanish as $y$ gets larger and larger. We remark that the condition of normalization below Eq.~\eqref{eta0lifsn0} restricts $z$ to the range $-1.2<z<0$ in the case $y_c\to\infty$. Finite values of $y_c$ leads to divergence at $y\geq y_c$ without adding restrictions to the values of $z$ due to the condition of normalization. In Fig.~\ref{fig7}, we display the Schr\"odinger-like potential \eqref{Uyn} for some values of the parameters with positive and negative $n$.

\begin{figure}[t!]
    \centering
    \includegraphics[width=6cm]{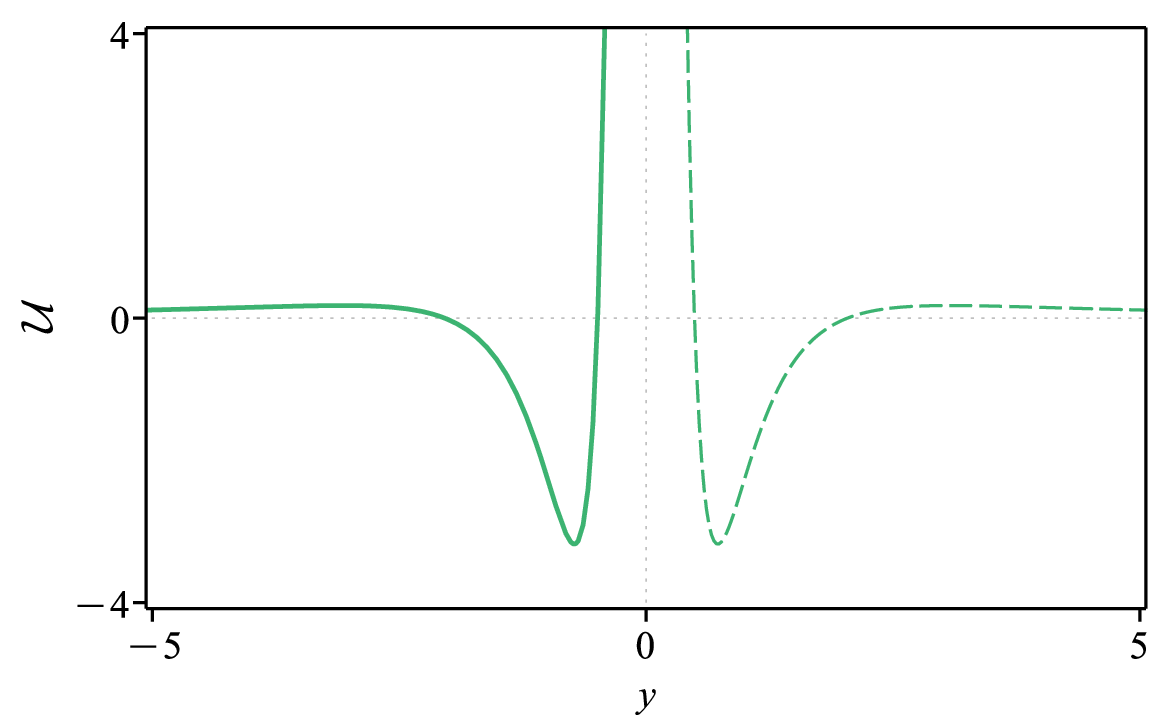}
    \caption{The Schr\"odinger-like potential in Eq.~\eqref{Uyn0} for $p=1$. The dashed line represents $\theta_c=1$ and $z=-1$, and the solid line stands for $\theta_c=3$ and $z=1$.}
    \label{fig8}
\end{figure}

Let us now analyze the case $n=0$. Before going further, we take into account that the case $\theta_c=0$ in three spatial dimensions does not allow for the existence of normalized zero modes (see the condition below \eqref{eta0lifsn0}). Thus, to work the case $n=0$ out, we take $\theta_c\neq0$. In this situation, we have $\theta_c=z+2$, so $z$ becomes the only parameter that modifies the model. Since the condition $\theta_c>2z$ must be satisfied, we must impose $z<2$ in $d=3$. We work with $p=1$, for which $z\in(-2,0)\cup(0,3/2)$. In this specific case, the behavior of $\Uc(y)$ both near the origin and asymptotically is $\Uc(y)\approx(16-z^2)/(4z^2y^2)$. So, at $y=0$, it diverges positively, and as $y\to\pm\infty$, it vanishes. The Schr\"odinger-like potential $\Uc(y)$ for $z=\pm1$ in Eq.~\eqref{Uyn0} is displayed in Fig.~\ref{fig8}, where we can see the symmetry $z\to-z$.

The number of modes depends on the value of $n$. With the chosen parameters, for positive values of $n$, the only surviving eigenstate of the Schr\"odinger-like equation is the zero mode. Negative values of $n$ lead to a richer analysis, as a potential with the form of an infinite well arises for finite values of $y_c$, so there are infinite bound states, while the limit $y_c\to\infty$ only supports the zero mode. For $n=0$, the only bound state is the zero mode. The absence of modes with negative eigenvalues ensure that the model is stable around small fluctuations.

\section{Conclusions}\label{secconclusions}
In this work, we have investigated the Maxwell-scalar model \eqref{LL} in radially symmetric spacetimes \eqref{ds2}, with the scalar and gauge fields coupled via the electric permittivity $\epsilon(\phi)$. To study this setup, we have developed a first-order framework based on the minimization of the energy compatible with the equations of motion. Interestingly, the procedure requires that the charge be exclusively given in terms of the factors of the line element.  Also, the first-order equation \eqref{fo} can be mapped into the corresponding one that arises in the canonical scalar field model in $(1,1)$ flat spacetime dimensions. This allowed us to obtain analytical results.

We then applied our method to some spacetimes. To get smooth localized structures, we have considered spacetimes with singularities at the origin in the $dt^2$ factor of the line element. First, we studied Tolman's metric VI \cite{tolman} in Eq.~\eqref{ds2tolman}, whose associated charge density presents a change of sign and the total charge is null. The scalar field solution connects two divergent values of the permittivity and engenders localized energy density. Its linear stability is described by a Schr\"odinger-like potential with the form of an infinite well whose zero mode is the ground state, ensuring stability. A similar analysis was done for the Exponential metric \eqref{ds2exp}, in which we have found positive charge density and finite charge, but a distinct profile for the solutions and energy densities, which can become compact with Schr\"odinger-like potential having the form of an infinite well. We remark that the limit $r_c\to\infty$ breaks the compact demeanor of the quantities: both the solution and energy density become extended and there is no infinite well in the Schr\"odinger-like potential anymore.

A third class that allows for the application of our method is the hyperscaling violating geometries, which is controlled by three parameters: $z$, $\theta_c$ and $d$. To apply our formalism, we have taken $\theta_c>2z$. In this situation, the charge density is non negative and the total charge is divergent. We have defined the parameter $n = \theta_c(d-1)/2-d-z+1$ that controls the solutions. For $n>0$, one must take $p\geq3$ to ensure that the minima of the potential are connected by the field configurations. The solutions are constant inside the limited space $r<r_c$ and engender a kink-like profile outside this interval, falling off with an exponential tail; its energy density presents a hole inside the limited space that is encircled by a radially-symmetric disk which vanishes asymptotically. For $n<0$, also with $p\geq3$, the solutions support a compact profile, with energy density that vanishes outside the compact space $r\leq r_c$. The case $n=0$ is special, as it allows us to take $p=1$, engendering a power-law solution with long range tail that is also present in the energy density. For all values of $n$, the energy is finite and is described by an analytical expression. We have investigated the stability to show the form of the Schr\"odinger-like potential and how it depends on $n$. Notwithstanding that, the model is linearly stable.

As perspectives, one may investigate the model \eqref{LL} on other geometric backgrounds. For instance, one may seek field configurations in which the change of variables \eqref{dx} maps $r$ into compact spaces. An example of this is the Reissner-Nordstr\"om metric in $(3,1)$ spacetime dimensions, with line element
\be\label{ds2rn}
ds^2 = f(r)\,dt^2 -\frac{1}{f(r)}\,dr^2 -r^2\,d\theta^2 -r^2\sin^2\theta\,d\vphi^2,
\ee
where $f(r)=1-2\mu/r+q^2/r^2$, and $\mu$ and $q$ are parameters that obey $q>\mu$ to avoid the presence of (coordinate) singularities in the metric outside the origin. 
As we have stated below Eq.~\eqref{ds2}, $f(r)$ must be non-negative in our formalism. Therefore, if one considers $q=0$, which recovers the Schwarzschild metric, one cannot use the first-order formalism in the form presented here because $f(r)$ would lead to a change of sign at $r=2\mu$ that would make $\Qc(r)$ become complex. An extension of the formalism presented here using coordinates adapted to the presence of horizons, such as Eddington-Finkelstein or Painlevé-Guldstrand coordinates, is thus necessary to address more general scenarios and will be explored elsewhere. It is also worth noting that the above metric is asymptotically flat, unlike the three examples considered in Sec.~\ref{secmodels}. This shows that asymptotic flatness is not needed to obtain localized solutions.

Within the first-order framework, Eqs.~\eqref{condbps} relate the charge density to the metric, so the charge must be calculated from Eq.~\eqref{Qbps}, which leads to finite charge, $Q=4\pi e$. The change of variables in Eq.~\eqref{dx} leads us to
\be\label{xRN}
x(r) = \frac{1}{\sqrt{q^2-\mu^2}}\left(\arctan\left(\frac{r-\mu}{\sqrt{q^2-\mu^2}}\right) -c_0\right),
\ee
where $c_0 = \arctan\left((r_c-\mu)/\sqrt{q^2-\mu^2}\right)$ was taken to make $x=0$ at $r=r_c$. The variable $x(r)$ is compact, obeying $x\in\left[0,\ell\right]$, where $\ell=\left(\pi/2-c_0\right)/\sqrt{q^2-\mu^2}$ is the width of the interval, such that $x=\ell$ for $r\to\infty$. This suggests that one must consider scalar field models in $(1,1)$ dimensions (governed by Eq.~\eqref{fox}) which support compact solutions. This type of solutions usually arise in the presence of V-shaped scalar potentials or in the limit in which the classical mass is infinite \cite{comp1,comp2,comp3}. The extremal case, $\mu\to q$, requires special care because it leads to a non-compact $x(r)$, but it also leads to gravitational BPS configurations  which are interesting on their own \cite{Hartle:1972ya}.

The spacetimes investigated in this work do not engender singularities outside the origin but do diverge at $r\to 0$ to impose the condition $\Qc(0)=0$. Consideration of regular spacetimes, with finite $f(r)$ as $r\to 0$ or restricted to domains with $r>0$ \cite{Alencar:2023wyf,Guerrero:2022msp,Simpson:2018tsi}, demands the consideration of Dirac delta contributions, which requires a careful extension of the approach presented here. An extension in that direction as well as in scenarios involving event horizons (coordinate singularities that affect the definition of $x(r)$), will be considered in a separate work. Other perspectives include the study of spacetimes with distinct features, such as the extensions of Tolman's metric VI to introduce anisotropy \cite{wittentolman} and/or Schwarzschild-like exteriors, or ones associated to black holes \cite{danilo3} and braneworlds \cite{rs2,sken,dewolfe,ksaki}. The case of braneworlds requires the inclusion of an extra spatial dimension of infinite extent, in a five-dimensional anti-de Sitter spacetime. This is similar to the case of the gauge/gravity duality, required to investigate strong interaction between quarks and gluons via holographic Einstein-Maxwell-scalar models \cite{GGD,Noronha}, and this brings more motivation to study braneworlds. In particular, we can think of braneworld scenarios with modified theories of gravity in the presence of non-constant curvature \cite{olmo1} and also, the addition of other scalar fields in a way similar to the recent work \cite{BL}, in which the presence of several scalar fields was used to control the internal structure to the brane. Another possibility of current interest concerns changing the Maxwell field to a non-Abelian set of vector field as it occurs, for instance, in the $SU(2)$ Yang-Mills case. A specific model of this kind was considered in Ref. \cite{su2}, and may stimulate new investigation in this line of research. We can also 
study spherically symmetric boson stars in Palatini $f(R)$ gravity, 
employing an approach that unveils a correspondence between modified gravity with scalar matter and general relativity with modified scalar matter \cite{olmo2}, using it in a way similar to the case recently considered in \cite{olmo3,Maso-Ferrando:2023nju}. We hope to return with new results on some of the above issues in the near future.

\acknowledgements{The authors acknowledge financial support from the Brazilian agencies Conselho Nacional de Desenvolvimento Cient\'ifico e Tecnol\'ogico (CNPq), grants Nos. 303469/2019-6 (DB), 402830/2023-7 (DB, MAM and RM), 306151/2022-7 (MAM) and 310994/2021-7 (RM), and Paraiba State Research Foundation (FAPESQ-PB) grants Nos. 2783/2023 (IA) and 0015/2019 (DB and MAM). This work is also supported by the Spanish Grants PID2020-116567GB-C21, PID2023-149560NB-C21 funded by MCIN/AEI/10.13039/501100011033, and by CEX2023-001292-S funded by MCIU/AEI.  The paper is based upon work from COST Action CaLISTA CA21109 supported by COST (European Cooperation in Science and Technology).
}

\end{document}